\documentclass[twoside,twocolumn,9pt]{article}
\usepackage{extsizes}
\usepackage[super,sort&compress,comma]{natbib}
\usepackage[version=3]{mhchem}
\usepackage[left=1.5cm, right=1.5cm, top=1.785cm, bottom=2.0cm, headheight=111pt]{geometry}
\usepackage{balance}
\usepackage{times,mathptmx}
\usepackage{sectsty}
\usepackage{graphicx}
\usepackage{lastpage}
\usepackage[format=plain,justification=justified,singlelinecheck=false,font={stretch=1.125,small,sf},labelfont=bf,labelsep=space]{caption}
\usepackage{float}
\usepackage{fancyhdr}
\usepackage{fnpos}
\usepackage[english]{babel}
\addto{\captionsenglish}{%
  \renewcommand{\refname}{References}
}
\usepackage{array}
\usepackage{droidsans}
\usepackage{charter}
\usepackage[T1]{fontenc}
\usepackage[usenames,dvipsnames]{xcolor}
\usepackage{setspace}
\usepackage[compact]{titlesec}
\usepackage{hyperref}

\definecolor{cream}{RGB}{222,217,201}

\begin{document}

\pagestyle{plain}
\fancyhf{}
\thispagestyle{plain}
\fancypagestyle{plain}{
\renewcommand{\headrulewidth}{0.0pt}
}

\makeFNbottom
\makeatletter
\renewcommand\LARGE{\@setfontsize\LARGE{15pt}{17}}
\renewcommand\Large{\@setfontsize\Large{12pt}{14}}
\renewcommand\large{\@setfontsize\large{10pt}{12}}
\renewcommand\footnotesize{\@setfontsize\footnotesize{7pt}{10}}
\makeatother

\renewcommand{\thefootnote}{\fnsymbol{footnote}}
\renewcommand\footnoterule{\vspace*{1pt}%
\color{cream}\hrule width 3.5in height 0.4pt \color{black}\vspace*{5pt}}
\setcounter{secnumdepth}{5}

\makeatletter
\renewcommand\@biblabel[1]{#1}
\renewcommand\@makefntext[1]%
{\noindent\makebox[0pt][r]{\@thefnmark\,}#1}
\makeatother
\renewcommand{\figurename}{\small{Fig.}~}
\sectionfont{\sffamily\Large}
\subsectionfont{\normalsize}
\subsubsectionfont{\bf}
\setstretch{1.125} 
\setlength{\skip\footins}{0.8cm}
\setlength{\footnotesep}{0.25cm}
\setlength{\jot}{10pt}
\titlespacing*{\section}{0pt}{4pt}{4pt}
\titlespacing*{\subsection}{0pt}{15pt}{1pt}

\fancyfoot{}
\fancyfoot[LO,RE]{\vspace{-7.1pt}\includegraphics[height=9pt]{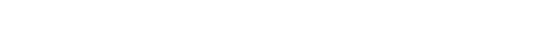}}
\fancyfoot[CO]{\vspace{-7.1pt}\hspace{13.2cm}\includegraphics{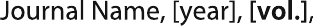}}
\fancyfoot[CE]{\vspace{-7.2pt}\hspace{-14.2cm}\includegraphics{head_foot/RF}}
\fancyfoot[RO]{\footnotesize{\sffamily{1--\pageref{LastPage} ~\textbar  \hspace{2pt}\thepage}}}
\fancyfoot[LE]{\footnotesize{\sffamily{\thepage~\textbar\hspace{3.45cm} 1--\pageref{LastPage}}}}
\fancyhead{}
\renewcommand{\headrulewidth}{0pt}
\renewcommand{\footrulewidth}{0pt}
\setlength{\arrayrulewidth}{1pt}
\setlength{\columnsep}{6.5mm}
\setlength\bibsep{1pt}

\makeatletter
\newlength{\figrulesep}
\setlength{\figrulesep}{0.5\textfloatsep}

\newcommand{\topfigrule}{\vspace*{-1pt}%
\noindent{\color{cream}\rule[-\figrulesep]{\columnwidth}{1.5pt}} }

\newcommand{\botfigrule}{\vspace*{-2pt}%
\noindent{\color{cream}\rule[\figrulesep]{\columnwidth}{1.5pt}} }

\newcommand{\dblfigrule}{\vspace*{-1pt}%
\noindent{\color{cream}\rule[-\figrulesep]{\textwidth}{1.5pt}} }

\makeatother
\twocolumn[
  \begin{@twocolumnfalse}
\vspace{3cm}
\sffamily
\begin{tabular}{m{4.5cm} p{13.5cm} }

\includegraphics{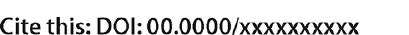} & \noindent\LARGE{\textbf{Chain ordering of phospholipids in membranes containing cholesterol: What matters?}} \\
\vspace{0.3cm} & \vspace{0.3cm} \\

 & \noindent\large{Fabian Keller,\textit{$^{a}$} and Andreas Heuer,\textit{$^{b}$}} \\

\includegraphics{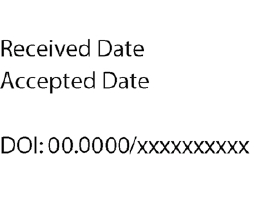} & \noindent\normalsize{Cholesterol (CHOL) drives lipid segregation and is thus a key player for the formation of lipid rafts and followingly for the ability of a cell to, e.g., enable selective agglomeration of proteins. The lipid segregation is driven by cholesterol's affinity for saturated lipids, which stands directly in relation to the ability of cholesterol to order the individual phospholipid (PL) acyl chains. In this work, Molecular Dynamics simulations of DPPC (Dipalmitoylphosphatidylcholine, saturated lipid) and DLiPC (Dilineoylphosphatidylcholine, unsaturated lipid) mixtures with cholesterol are used to elucidate the underlying mechanisms of the cholesterol ordering effect. To this end, all enthalpic contributions, experienced by the PL molecules, are recorded as a function of the PL's acyl chain order. This involves, the PL-PL, the PL-cholesterol interaction, the interaction of the PLs with water, and the interleaflet interaction. This systematic analysis allows one to unravel differences of saturated and unsaturated lipids in terms of the different interaction factors. It turns out that cholesterol's impact on chain ordering stems not only from direct interactions with the PLs but is also indirectly present in the other energy contributions.
Furthermore, the analysis sheds light on the relevance of the entropic contributions, related to the degrees of freedom of the acyl chain.} \\

\end{tabular}

 \end{@twocolumnfalse} \vspace{0.6cm}
  ]


\renewcommand*\rmdefault{bch}\normalfont\upshape
\rmfamily
\section*{}
\vspace{-1cm}


\footnotetext{\textit{$^{a,b}$Institute of Physical Chemistry, University of Münster, Corrensstra{\ss}e 28, 48149 Münster, Germany. Fax: +49 251 83 29159 Tel: +49 251 83-29177; E-mail: andheuer@uni-muenster.de}}



\section{Introduction}

\begin{figure*}
    \centering
    \includegraphics[width=16cm]{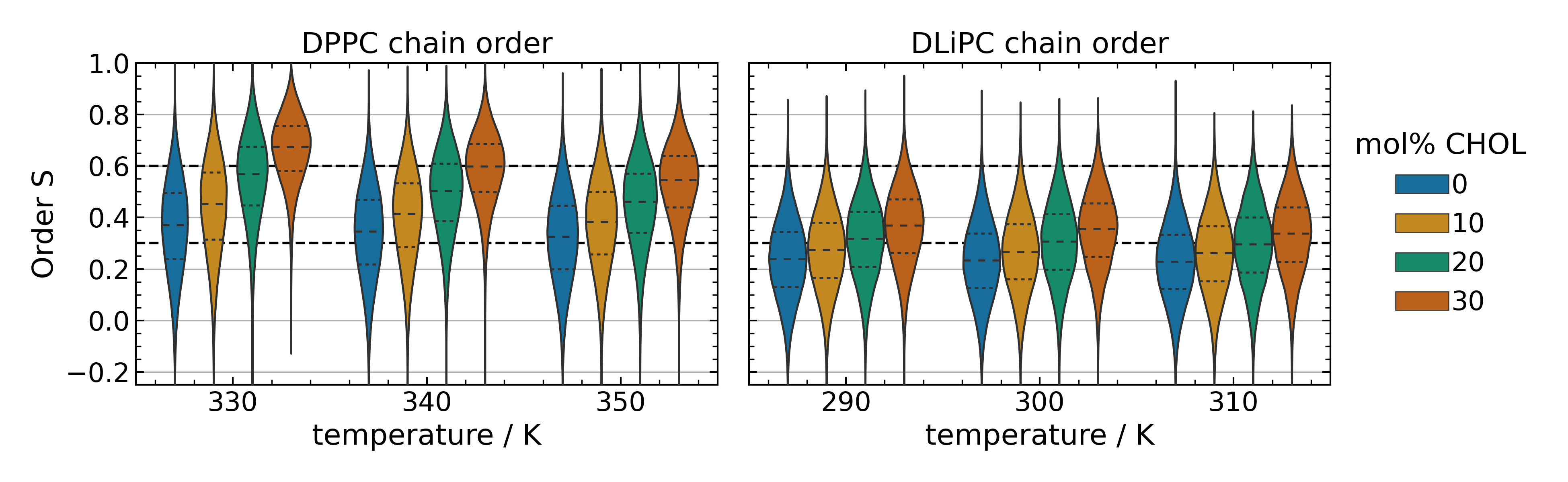}
    \caption{ DPPC (left) and DLiPC (right) chain order parameter distribution in simulations varying in temperature and cholesterol concentration.}
    \label{fig:pscd}
\end{figure*}

The functionality of Plasma Membranes (PM) of higher eukaryotic cells is controlled by a complex interplay of lipids of varying types, proteins, and other bioactive compounds. Within this interaction network, cholesterol is one of the most important ingredients exhibiting outstanding features as it is able to tune the mechanical properties and phase behavior of lipid bilayers.
Ranging from concentrations of less than 10 mol\% in organelle membranes to concentrations of 20-30 mol\% of cholesterol in most of the PM, cholesterol leads to tighter lipid packing, making bilayers less permeable for smaller molecules and thus enhancing the PMs ability to shield the cytosol from the outside world.\cite{Sackmann1995, Sezgin2017, Levental2016, Rog2014, Simons2011, Heberle2011, Fan2010, Rog2009, Meer2008}

The formation of rafts is said to be directly linked to cholesterol's affinity towards high-melting (saturated) lipids and thus has been well discussed in the literature.
The idea that cholesterol's effect on bilayer structure is universal and thus independent of composition was replaced by the notion that cholesterol's affinity to a lipid decreases with the lipids unsaturation and varies with different head groups.\cite{Pan2008, Pan2009, Engberg2016, Yang2016} A recent study closer investigated the effects of cholesterol on unsaturated lipid bilayers, which were assumed to be unaffected by cholesterol by any means.\cite{Chakraborty2020} In this study, they indeed found a, though weaker, ordering effect of cholesterol on unsaturated lipids.

Regen et al. used a technique coined nearest-neighbor recognition (NNR), in which pairs of like and unlike molecules are counted and hence a free energy of interaction can be calculated from it. In one approach, they compared kinked structures of cholesterol to compare the so-called template effect, where the rigid body of cholesterol is said to act as a template for ordering lipid acyl chains and the proposed umbrella model\cite{Huang1999}, where PLs act as an umbrella, shielding the cholesterol body from unfavorable water interactions.\cite{Daly2011, Krause2014} Their results point towards the template mechanism for the description of cholesterols' ordering effect. In another study, they compared kinked and unkinked modifications of PLs to directly identify favorable interactions with saturated lipids vs unfavorable interactions with unsaturated lipids\cite{Mukai2017} and via a Monte Carlo model, derived from NNR results, propose that this affinity is driven mainly by enthalpy and not by entropy.\cite{Wang2018}

The important feature of domain registration is discussed to be highly dependent on composition.\cite{Polley2014} The degree of the relation of leaflet interdigitation on domain registration is discussed contradictory. While in a recent study the interdigitation is identified as a key driving force for domain registration\cite{Seo2020} another work negated any major relevance\cite{Tian2016}. In any way, leaflet interdigitation plays a role in ordering lipids. The higher the degree of unsaturation, the higher the leaflet interdigitation\cite{Zhang2019}. Additionally, it is reported that cholesterol decreases leaflet interdigitation.\cite{Leeb2018}

Cholesterol is known to decrease the water permeation by increasing acyl chain order. Even though the permeation is lowered and cholesterol does not have an impact on the PL head group region\cite{McIntosh1978}, studies have shown that the hydrophobicity of the polar head group region is decreased.\cite{Subczynski1994} An early MD work indicates increased hydration of the head region.\cite{Pasenkiewicz2000} These findings are strongly related to the aforementioned umbrella model, which relates cholesterol's ability to order acyl chains and properties at the bilayer-water interface.

All these examples show the complex nature of cholesterol's impact on PL ordering. In this work we are taking a systematic look into each of the different sites cholesterol takes effect on from an energetic point of view. Specifically, we elucidate cholesterol's effect on PL-PL interactions, PL-water interactions, interleaflet interactions, and, of course, the direct PL-cholesterol interaction. Each of these shed light on the interplay between those different effects.
For this purpose, we employed thorough Molecular Dynamics simulations of binary mixtures of DPPC (Dipalmitoylphosphatidylcholine) and DLiPC (Dilineoylphosphatidylcholine) with cholesterol, respectively, with concentrations ranging from pure PL systems to systems containing 30\% cholesterol. The choice of a fully saturated (DPPC) and a polyunsaturated (DLiPC) lipid, even though rarely occurring in living organisms, provide the two extreme cases of cholesterol's mode of action. We expect the behavior of cholesterol to change significantly between these limits. We used the well established CHARMM36 force-field, which has proven to very well reproduce lipid membrane properties and, in particular, acyl chain order.\cite{Pluhackova2016,Rosales2020} The resulting order parameter distributions for the different bilayer compositions simulated at different temperatures are shown in figure \ref{fig:pscd}. In accordance with the expected properties of cholesterol, it shows a distinctively higher ordering effect on DPPC than on DLiPC. Which interactions are needed to correctly describe cholesterol ordering of saturated and unsaturated lipids? This is the key question that guides the analysis of this work.

\section{Methods}

\begin{figure}
    \centering
    \includegraphics[width=8cm]{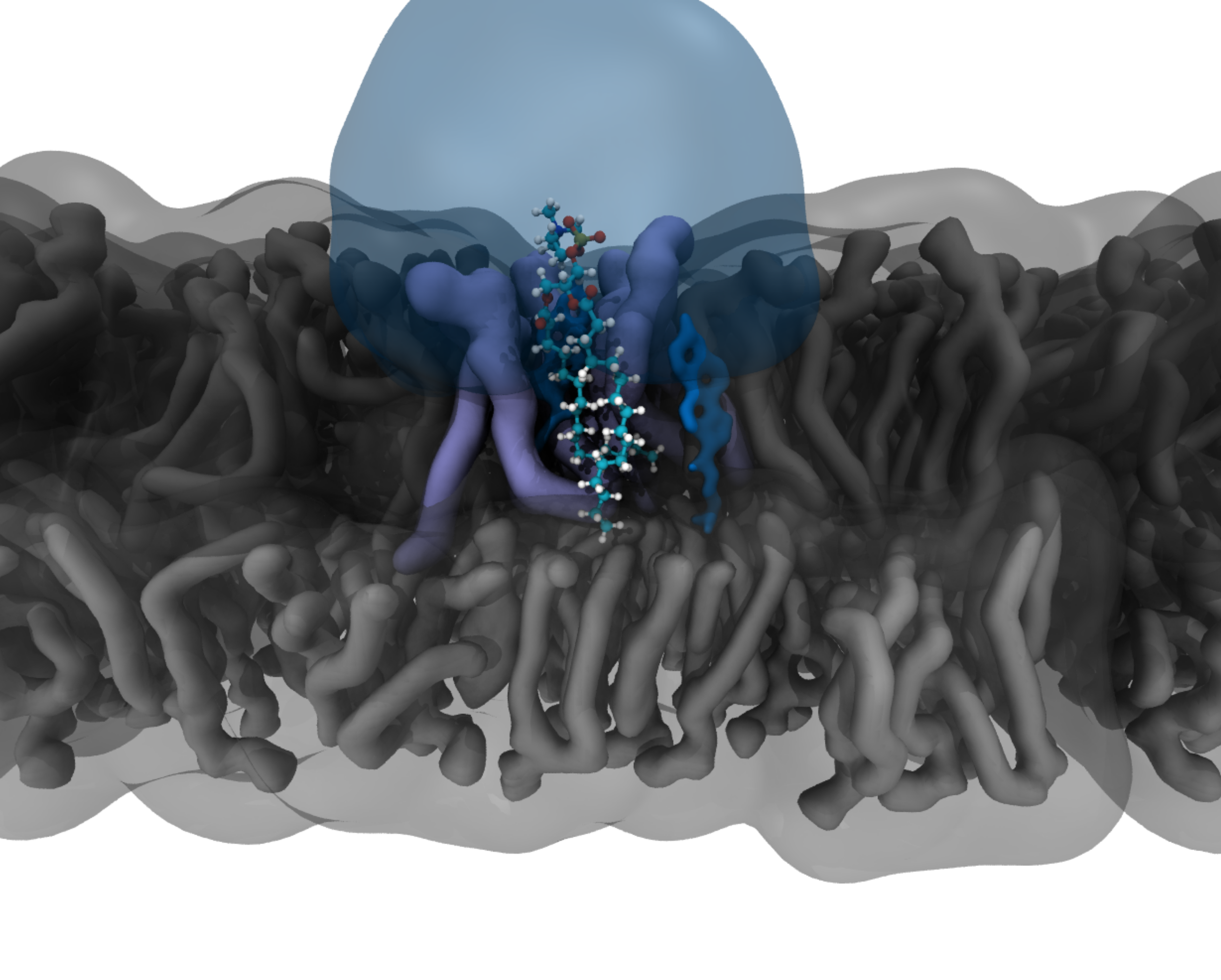}
    \caption{
        Identification of the different types of interaction for a PL (as CPK): nearest PL neighbors (purple), nearest CHOL neighbors (blue), non-nearest $N_{out}$ neighbors (grey), and the surrounding water (blue shading).}
    \label{fig:int_comp}
\end{figure}

\subsection{Molecular dynamics simulations}
The bilayer structures were prepared using the CHARMM-GUI Web-based graphical interface.\cite{Lee2016} All Molecular dynamics (MD) simulations were conducted using the CHARMM36 force-field\cite{Klauda2010a} and Gromacs~2018 MD software package.\cite{Abraham2015} The systems were equilibrated using the established CHARMM-GUI parameter set. The TIP3P model was used as water model.\cite{Jorgensen1983} To maintain the temperature the Nos{\'{e}}-Hoover algorithm was used with a coupling constant of 1~ps, coupling bilayer and solvent separately.\cite{Nose1984} To maintain the pressure at 1~atm the Parrinello-Rahman barostat was used with a coupling constant of 5~ps and a compressibility of 4.5x10$^{-5}$~bar$^{-1}$ \cite{Parrinello1981}. Hydrogen bonds were constraint using LINCS.\cite{Hess1997} Particle mesh Ewald electrostatics were used with a cutoff of 1.2~nm.\cite{Darden1993} The Lennard-Jones potential was shifted to zero between 1.0 and 1.2~nm, and a cutoff of 1.2~nm was used. The nonbonded interaction neighbor list was updated every 20 steps, using a cutoff of 1.2~nm. The temperature ranges, trajectory lengths, and compositions for each bilayer simulation are shown and summarized in table \ref{tab:MDparam} in the supporting information.

\subsection{Derivation of measured parameters}
All analysis heavily relied on the MDAnalysis package for Python.\cite{Agrawal2011,Gowers2016} All visualizations of the simulations were created using VMD.\cite{Humphrey1996}
\paragraph*{The lipid chain order parameter S} is defined as $S = \left< 1.5\cos^2\theta - 0.5 \right> $, where $\theta$ denotes the angle between the vector spanned by every second carbon atom in a lipid chain and the average tilt vector per leaflet. Brackets indicate an average value over both chains and their carbon vectors. This definition of calculating the order parameter differs from the commonly used S$_\mathrm{CH}$ order parameter, as we do not consider hydrogen atoms in this calculation, thus our definition is shifted by $S = \left< 1.5\cos^2 \pi/2 - 0.5 \right> = -0.5$ and thus S=-0.5 being in the lowest order state, while S=1.0 being the highest order state (linear chains). This has the advantage of easier comparability between the parameters of saturated and of unsaturated PLs.
\paragraph*{Radial distribution functions} (RDF) were calculated using gromacs analysis tools.\cite{Abraham2015} All RDFs were calculated individually for each leaflet using in-plane distances and were averaged. For the PL molecules and for CHOL the phosphor and the hydroxyl oxygen atoms respectively were used as reference positions.
\paragraph*{The overlap score} is the average overlap integral of groups of lipids based on the Bhattacharyya distance defined as $BC_i(p,q) = \sum\limits_{z \in Z} \sqrt{p_i(z) \cdot q_i(z)} $ where $p_i(z)$ and $q_i(z)$ are the normalized distributions of the bilayer height (z-component) of the positions of the last 3 carbon atoms for a lipid group $i$. In this manner, each PL molecule was assigned a group of neighboring PLs, with their P atoms located within 1~nm distance of the respective PL molecule's P atom. Note that this definition is analogous to the nearest-neighbor $N_N$ definition. The $BC$ was then calculated for the z-position distributions of the lipid group ($p_i(z)$) and the lipids within the same radius but in the opposing leaflet ($q_i(z)$). In this definition, a complete overlap of the last carbons in the PL chains would lead to an overlap score of 1. To relate the overlap to the order parameter, the order parameter of the central lipid within each group was used.

 \subsection{Determination of enthalpic contributions}
 We separated the environment of a PL into different contributions to the ordering effect of PL chains, being interactions with its nearest neighbors (N$_n$), interactions with lipids of the outer neighboring shells (black region, $N_{out}$), the PL interactions with the opposing leaflet (white region) and the interaction with the surrounding water (blue region) (figure \ref{fig:int_comp}) and calculate the average sum of all interaction energy contributions of a PL with its surroundings as a function of the PL's chain order. The interaction functions are averaged over simulations at different temperatures where the bilayers are found in an Ld/Lo state (290~K to 330~K for DLiPC, 330~K to 350~K for DPPC). Each energy is an average within an order parameter range of 0.1 and its standard deviation in each range lies at roughly 90~kJ/mol. The standard deviation is similar in all order parameter ranges, for both DPPC and DLiPC and in all CHOL compositions. Importantly, all averages presented in this work were determined with a numerical uncertainty of the mean smaller than 1~kJ/mol.

\section{Results and discussion}
\begin{figure*}
    \centering
    \includegraphics[width=16cm]{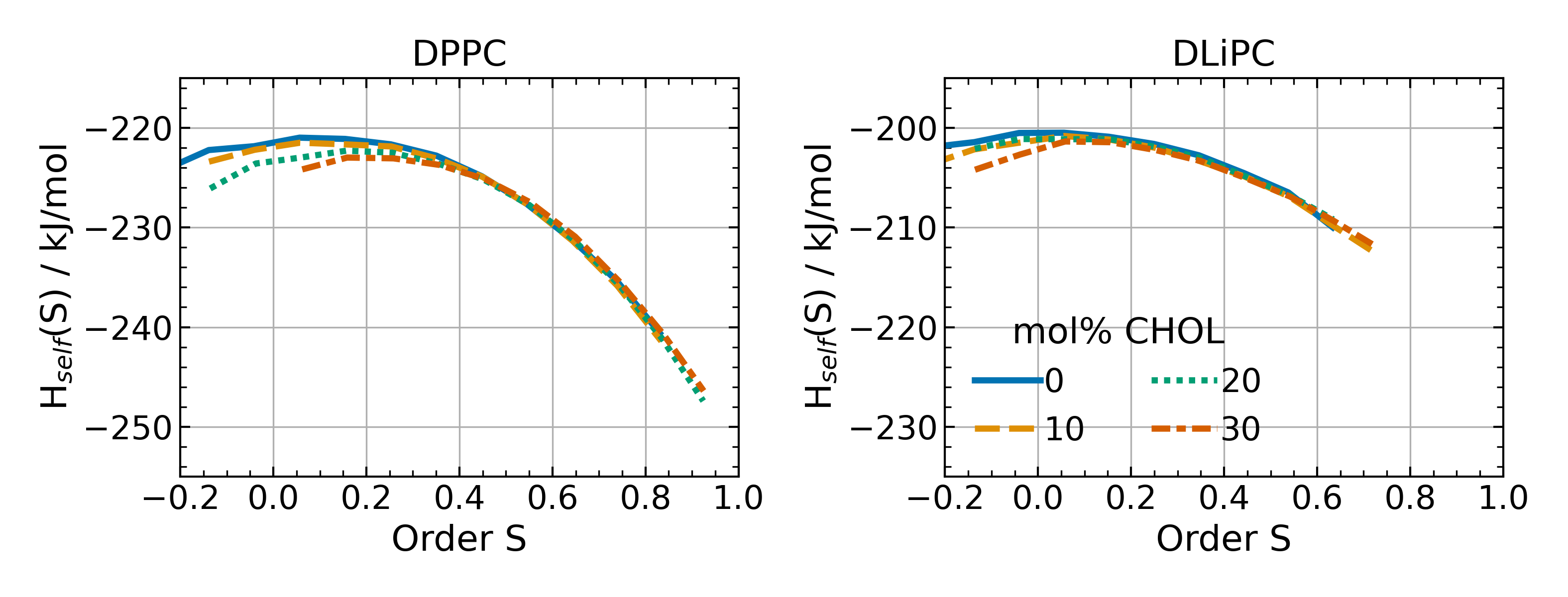}
    \caption{Self-energies of DPPC (left) and DLiPC (right) in PL/CHOL mixtures of varying CHOL concentrations. Here the analysis is restricted to the mutual interactions of atoms of a single lipid.}
    \label{fig:Eself}
\end{figure*}

\subsection{PL-PL interaction}
We start with the self-energy of a randomly chosen PL with order parameter S, determined as the average over its two acyl chains. For both PL the self-energy profiles exhibit an overall similar shape (figure \ref{fig:Eself}) in the range of order parameters, accessible for both systems. Furthermore, even at the highest CHOL concentration no impact of CHOL is visible. Naturally, the interaction of the two acyl-chains dominates the dependence on the order parameter (data not shown).  Note that for order parameters smaller than 0.4 the S-dependence is very weak.

When analyzing the interaction of a randomly chosen PL molecule with the remaining PL molecules, we distinguish nearest neighbor N$_N$ interaction from the interaction with outer neighbor shells N$_{out}$. They are separated by a distance cutoff taken from the position of the second minimum of the respective radial distribution functions with PL-phosphor and CHOL-hydroxy oxygen atoms as reference positions (figure \ref{fig:rdf_PLPL} and \ref{fig:rdfPLO}).

\begin{figure*}
    \centering
      \includegraphics[width=16cm]{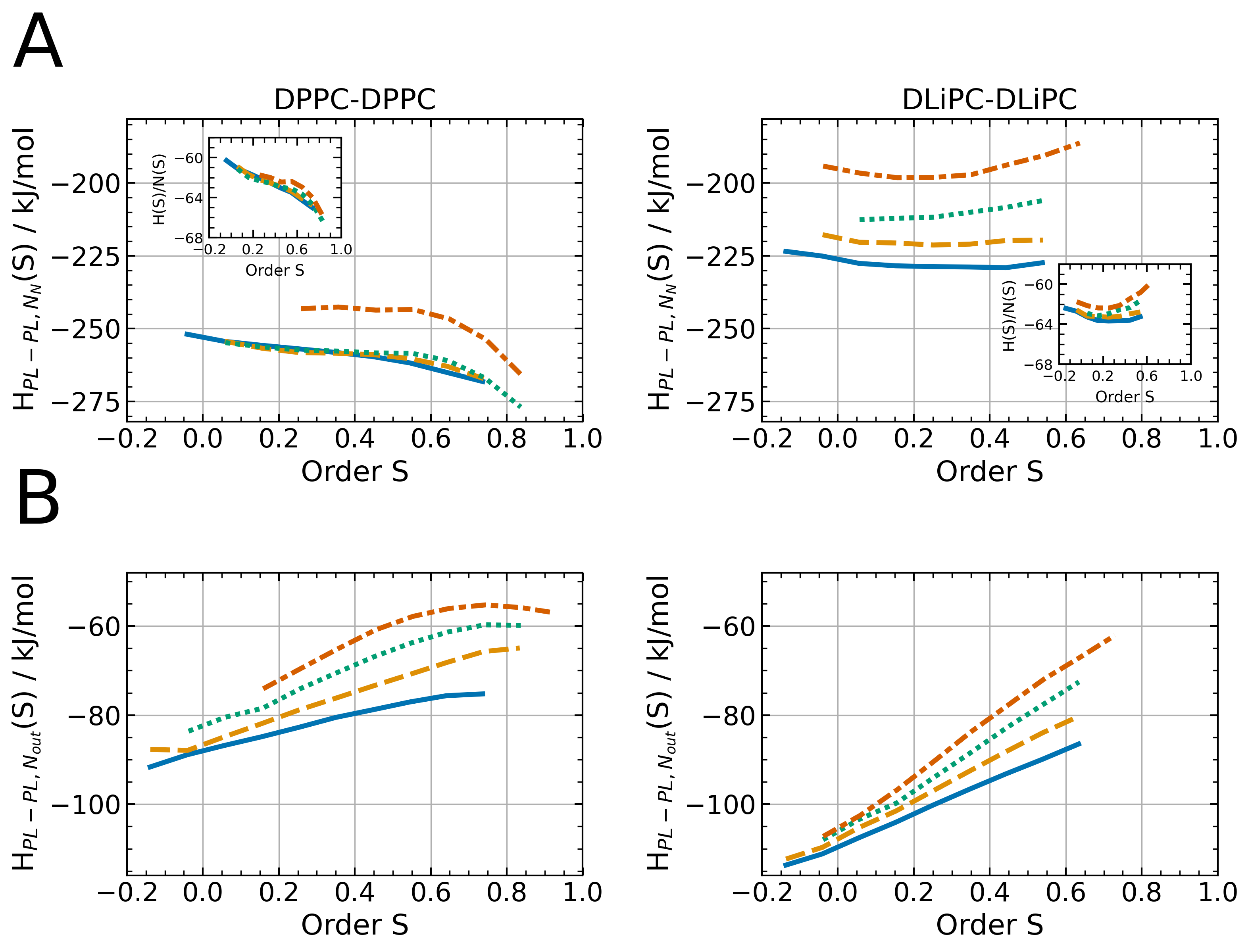}
    \caption{A) Nearest neighbor N$_N$ PL interaction as a function of PL order. B) interaction of a PL with the outer neighbor shell N$_{out}$. The interaction functions are an average from simulations of the respective PL/CHOL mixture for temperatures of 290 to 320~K (DLiPC) and 330 to 350~K (DPPC). Solid blue, dashed yellow, dotted green and dot-dashed red indicate that energies were derived from bilayers with CHOL concentrations of 0, 10, 20 and 30 mol\% respectively.}
    \label{fig:Eplpl}
\end{figure*}

The N$_N$ interaction energies as a function of the order parameter are shown in figure~\ref{fig:Eplpl} both for DPPC and DLiPC. To interpret this behavior more closely we have also determined the average number of PL neighbors (see figure \ref{fig:Nneibs} in the SI). Now a few important conclusions can be drawn. First, after dividing the interaction energy by the average number of neighbors the dependence on the CHOL concentration basically disappears for both PL (figure~\ref{fig:Eplpl} inset). Thus, the energy per PL-PL pair for given order parameter is independent of the CHOL concentration.  Second, the normalized energy displays a significant S-dependence mainly for DPPC, i.e. favoring the interaction with elongated acyl chains. Third, whereas for DPPC the number of nearest neighbors does not depend on the CHOL concentration (except for the 30\% data), a strong dependence is observed for DLiPC. This observation reflects the well-known condensing effect.\cite{Daly2011} It states that for saturated lipids CHOL may additionally be integrated into the membrane without changing the PL-PL neighborhood. In contrast, for unsaturated lipids this integration is slightly weaker. For example, upon adding 20\% CHOL the number of nearest neighbors is reduced by approx. 7\%.

As expected the interaction from the outer neighbor shells N$_{out}$ is much smaller. However, due to its dependence on the order parameter and on the CHOL concentration, it cannot be neglected for a complete description of the enthalpic contributions in a membrane. The DPPC-DPPC interaction somewhat depends on the CHOL concentration even for small CHOL concentrations. This is in contrast to the N$_N$ interaction. Furthermore, the decrease of the interaction strength with increasing order parameter is opposite to the N$_N$ interactions. These observations are a consequence of the fact that interaction beyond the nearest neighborhood is particularly strong when both interacting lipids have a disordered acyl chain because of the increased probability of their encounter. This explains why for low order parameters of the central PL and low CHOL concentration,  implying overall lower order parameters in the membrane, the interaction is strongest.

\subsection{PL-CHOL interaction}

\begin{figure*}
    \centering
    \includegraphics[width=16cm]{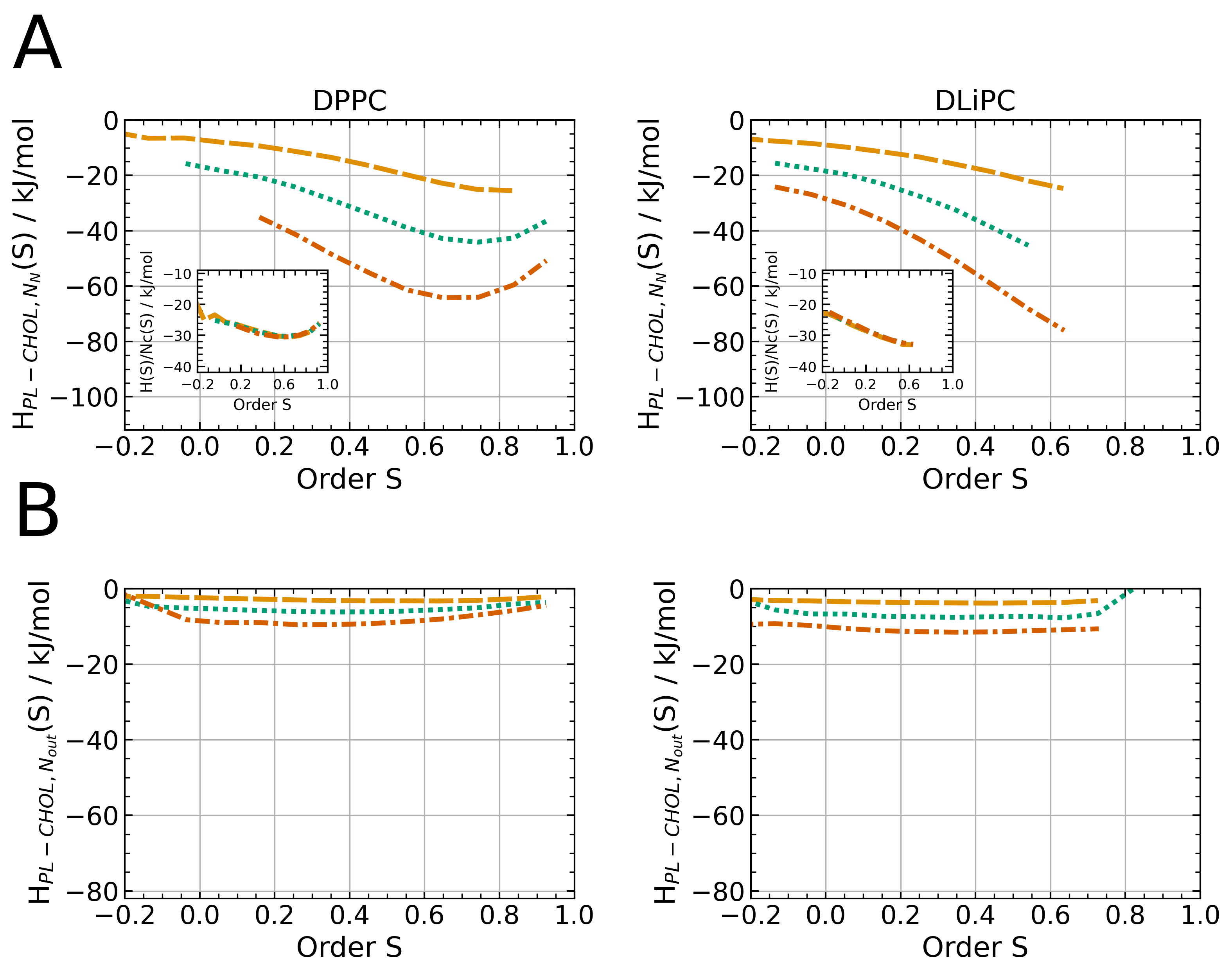}
    \caption{A) Nearest neighbor N$_N$ and B) outer neighbor N$_{out}$ PL-CHOL interaction as a function of PL order. The interaction functions are an average from simulations of the respective PL/CHOL mixture for temperatures of 290 to 320~K (DLiPC) and 330 to 350~K (DPPC). Solid blue, dashed yellow, dotted green and dot-dashed red indicate that energies were derived from bilayers with CHOL concentrations of 0, 10, 20 and 30 mol\% respectively.}
    \label{fig:Eplc}
\end{figure*}

One may expect that the PL-CHOL interaction is essential to understand the different impact of CHOL on the ordering of both lipids. The results for the DPPC-CHOL and DLiPC-CHOL N$_N$ interactions are shown in figure~\ref{fig:Eplc}, A. Naturally, for higher CHOL concentration a stronger interaction is observed. This can be expressed even more quantitatively. It turns out that after normalization by the actual number of surrounding CHOL-molecules (for a given order parameter) the interaction energy per PL-CHOL pair does no longer depend on the number of surrounding CHOL-molecules (inset of figure \ref{fig:Eplc}). Stated differently, the PL-CHOL interaction is simply additive even for larger concentrations where a PL with, e.g., an order parameter around 0.5 is surrounded by approx. 2 CHOL-molecules.

Closer inspection shows that the DPPC-CHOL and DLiPC-CHOL N$_N$ interactions, both in absolute and relative terms,  are very similar for order parameters up to 0.5 (figure~\ref{fig:Eplc}, A). Thus, for disordered chains the PL-CHOL interaction gives rise to an increase of PL order, in particular for higher CHOL concentrations. On a qualitative level this is compatible with the increase of the average order parameters of the system with increasing CHOL concentration.

For high order parameters the DPPC-CHOL and DLiPC-CHOL interactions start to differ. Only the DPPC-CHOL N$_N$ interaction strength exhibits a maximum (energy minimum) which is found at S=0.7. This implies the presence of an optimum configuration of a saturated acyl chain, adapting to CHOL's rigid body. This observation also reflects the unique ability of CHOL to inhibit gel formation for DPPC and supports the idea by Regen et al.\cite{Daly2011} that the CHOL body acts as a template for chain ordering and that any change of its structure will lead to a drastically different ordering capability. Interestingly, when considering the interaction energy, normalized by the number of CHOL-neighbors, also DLiPC approaches such a maximum so that in analogy to DPPC there also exists an optimum interaction motif. Of course, due to the presence of double bonds in the acyl chains, order parameters significantly beyond that maximum cannot be explored.

When comparing the energy gain $\Delta E$ for ordering chains from S=0.3 to S=0.5 between the PL-PL and the PL-CHOL action normalized to the number of the respective PL and CHOL neighbors, the importance of the PL-CHOL interaction for the overall ordering can be understood energetically, e.g. at 20\% CHOL the energy gain of the PL-PL interaction lies at $\Delta E_{PL-PL}=-0.55$~kJ/mol, while it is as high as $\Delta E_{PL-CHOL} = -1.35$ for the PL-CHOL interaction and thus being roughly 2 to 3 times larger.

In contrast to the PL-PL interaction, the  PL-CHOL N$_{out}$ interaction has only a very small contribution to the overall energy of chain ordering (figure~\ref{fig:Eplc}, B).

\subsection{Interleaflet interaction}

\begin{figure*}
    \centering
    \includegraphics[width=16cm]{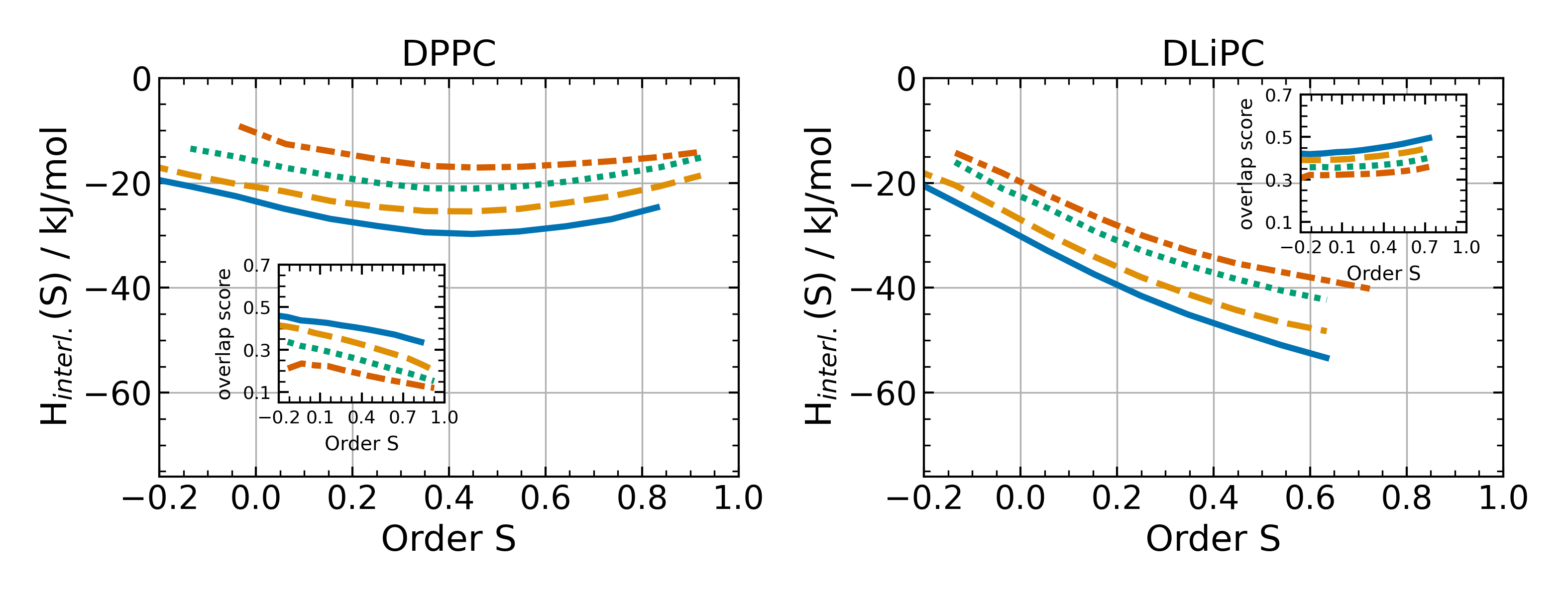}
    \caption{Interaction of PLs with lipids of the opposing leaflet as a function of the PL chain order parameter. The insets show the relation between a lipids overlap score, defined as the correlation of the position distributions of the last three carbon atoms, with its chain order parameter for DPPC and DLiPC bilayers. Solid blue, dashed yellow, dotted green and dot-dashed red indicate that energies were derived from bilayers with CHOL concentrations of 0, 10, 20 and 30 mol\% respectively.}
    \label{fig:Einterleaf}
\end{figure*}

Domain registration requires an interaction between the opposing leaflets of a bilayer. Studying the dependence on the order parameter of the randomly chosen PL, DPPC displays a non-monotonic S-dependence in contrast to DLiPC with a monotonous increase of the interleaflet interaction strength with order parameter (figure~\ref{fig:Einterleaf}). In both cases, the addition of CHOL reduces the interleaflet interaction at a given order parameter.

Structurally, the interleaflet interaction can be characterized by the overlap score (see Methods). Anticipating a dependence on that overlap score, we calculated it as a function of the PLs order parameter (insets of figure~\ref{fig:Einterleaf}).
Lin et al. found that the higher the lipid unsaturation, the higher the leaflet interdigitation\cite{Zhang2019}, and, indeed, we find that the overall interdigitation is higher for DLiPC. While chain ordering of DPPC leads to a decrease in DPPC chain intercalation, the DLiPC intercalation even weakly increases with increasing DLiPC order parameter. In accordance with the work of Leeb and Maibaum, the overlap score decreases for both lipids with increasing CHOL concentration.\cite{Leeb2018} Additionally, we have calculated the interleaflet interaction as a function of order parameter {\it and} overlap score (figure \ref{fig:overlapmaps} in the SI). For both PL the interaction energy becomes stonger if either the order parameter or the overlap score increases, albeit the dependence on order parameter (for fixed overlap score) is even stronger for DLiPC whereas for DPPC it is only relevant for small order parameters.

These observations allow for a simple interpretation of the results in figure~\ref{fig:Einterleaf}. For DLiPC, both the direct effect of the increasing order parameter as well as the additional implicit effect of the increasing overlap score promote an increase of the interleaflet interaction strength. In contrast, for DPPC two opposing effects have to be taken into account. Starting from low order parameters, an increase of S gives rise to a stronger interleaflet interaction. However, due to the decrease of the overlap score with order parameter, the interaction strength decreases for larger S, where the direct S-dependence vanishes (figure \ref{fig:overlapmaps} in the SI).

As again shown in figure \ref{fig:overlapmaps} the addition of CHOL decreases the interleaflet interaction strength even for fixed order parameter and fixed overlap score. This observation, reflecting some complex impact of CHOL on the different interaction terms, naturally explains the dependence on the CHOL concentration for DPPC and DLiPC in figure~\ref{fig:Einterleaf}.

The influence of leaflet intercalation on domain registration is discussed contradictory by Shinoda et al. and Cheng et al.\cite{Seo2020,Tian2016} The former found a high influence of leaflet intercalation simulating different asymmetric SM/DOPC/CHOL bilayer compositions, while the latter found no influence of leaflet intercalation in POPC/DSPC/CHOL bilayers. It is reasonable to assume that the presence of domain registration goes along with a strong interleaflet interaction which, naturally, will depend on the local composition. In the SM/DOPC/CHOL setup a clear phase separation is visible with regions enriched in the unsaturated DOPC and depleted of CHOL. As concluded from our results, those regions exhibit a strong interleaflet interaction, as all three factors apply here (inherent disorder of chains of DOPC, increased intercalation and depletion of CHOL). On the other hand, the POPC/DSPC/CHOL system should show a more or less homogenous lateral distribution and thus, even though there is an overlap, should have a significantly weaker interleaflet interaction than is the case for the regions enriched in DOPC.

\subsection{PL-water interaction}

\begin{figure*}
    \centering
    \includegraphics[width=16cm]{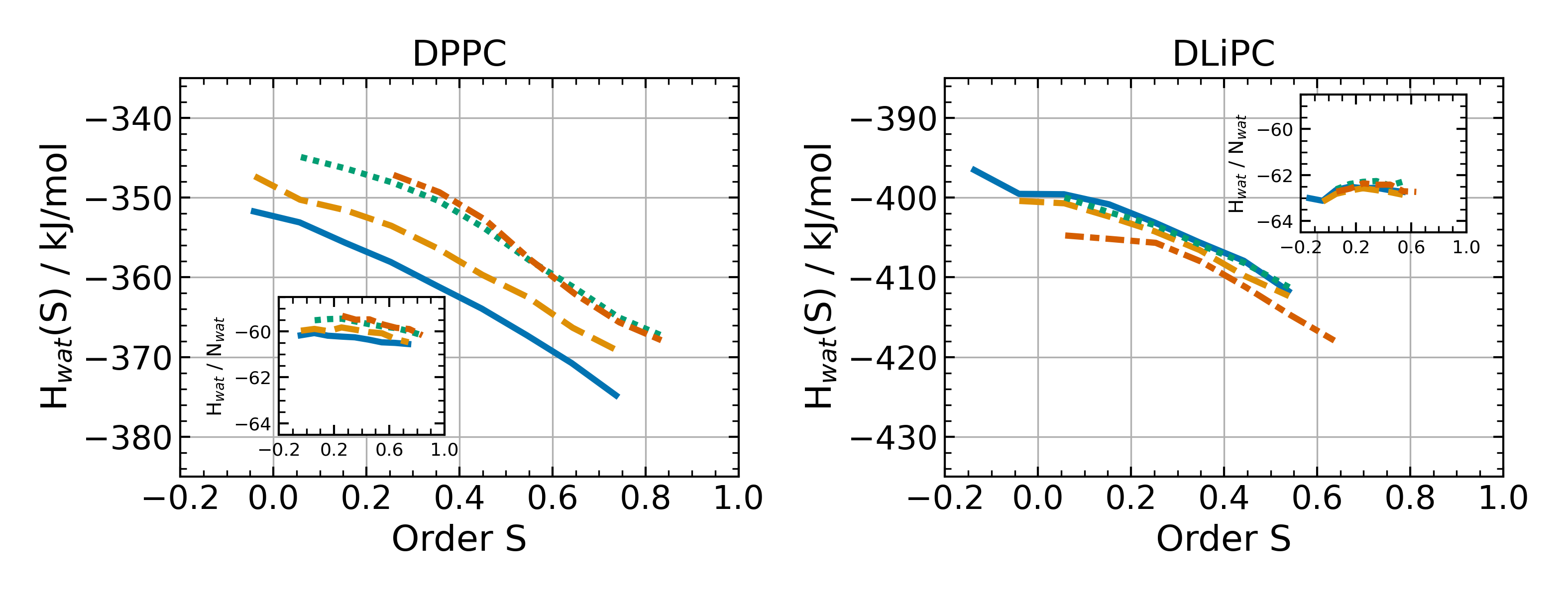}
    \caption{Interaction of a PL with water as a function of the PLs chain order parameter. The inset shows the ratio of water interaction and the respective average number of water molecules, thus constituting a PL-water pair energy. Solid blue, dashed yellow, dotted green and dot-dashed red indicate that energies were derived from bilayers with CHOL concentrations of 0, 10, 20 and 30 mol\% respectively.}
    \label{fig:Ewat}
\end{figure*}

The effect of CHOL on the bilayer water interfacial structure has been widely discussed and can be conflated to the so-called umbrella model\cite{Huang1999}, proposing an unfavorable interaction of water with the CHOL body and, accordingly, the PL head groups acting as an umbrella to shield CHOL from water. In this manner, increased CHOL concentration should lead to an increase in the interaction of PLs with water. Furthermore, as discussed in the introduction, CHOL influences the PL head group hydration.\cite{Subczynski1994, Pasenkiewicz2000}

We first looked at the effect of order parameter and CHOL concentration on the hydration of the PL head group region (figure~\ref{fig:watneibs} in the SI). A neighbor cutoff was determined from the position of the first minimum of the PL-OH2 RDFs (figure~\ref{fig:rdfh2o}).
Surprisingly, both for DPPC and DLiPC the number of neighboring water molecules around the PL phosphor atom increases with order, though the hydration of the head group region is larger for DLiPC. Cholesterol has an opposite, albeit weak,  effect on the two lipid types: While for DPPC the average number of water neighbors decreases with increasing CHOL concentration, it increases for DLiPC.
This change of the number of water molecules with CHOL concentration is stronger for DPPC and, consequently, the trend is visible in the water interaction energies as well (figure~\ref{fig:Ewat}).
To distinguish between the effects of the merely increased number of water molecules that interact with the PL and an optimization of a water-PL interaction configuration, we calculated the energy gain per water molecule in the first neighbor shell around the PL phosphor atoms (inset of figure~\ref{fig:Ewat}). Remarkably, the energy contribution per water molecule no longer depends on the order of the PL nor on the CHOL concentration.

Thus, the dependence of the water interaction on lipid chain order parameter  and CHOL concentration is solely a consequence of the slightly varying degree of hydration. We, therefore, found no evidence for a mechanistic effect of CHOL on the lipid water interaction as proposed by the umbrella model and agree with the literature in that CHOL does not affect the PL head group region, measured by the lipid interaction profile.

\subsection{Overall perspective on the enthalpic contributions}

\begin{figure*}
	\centering
	\includegraphics[width=16cm]{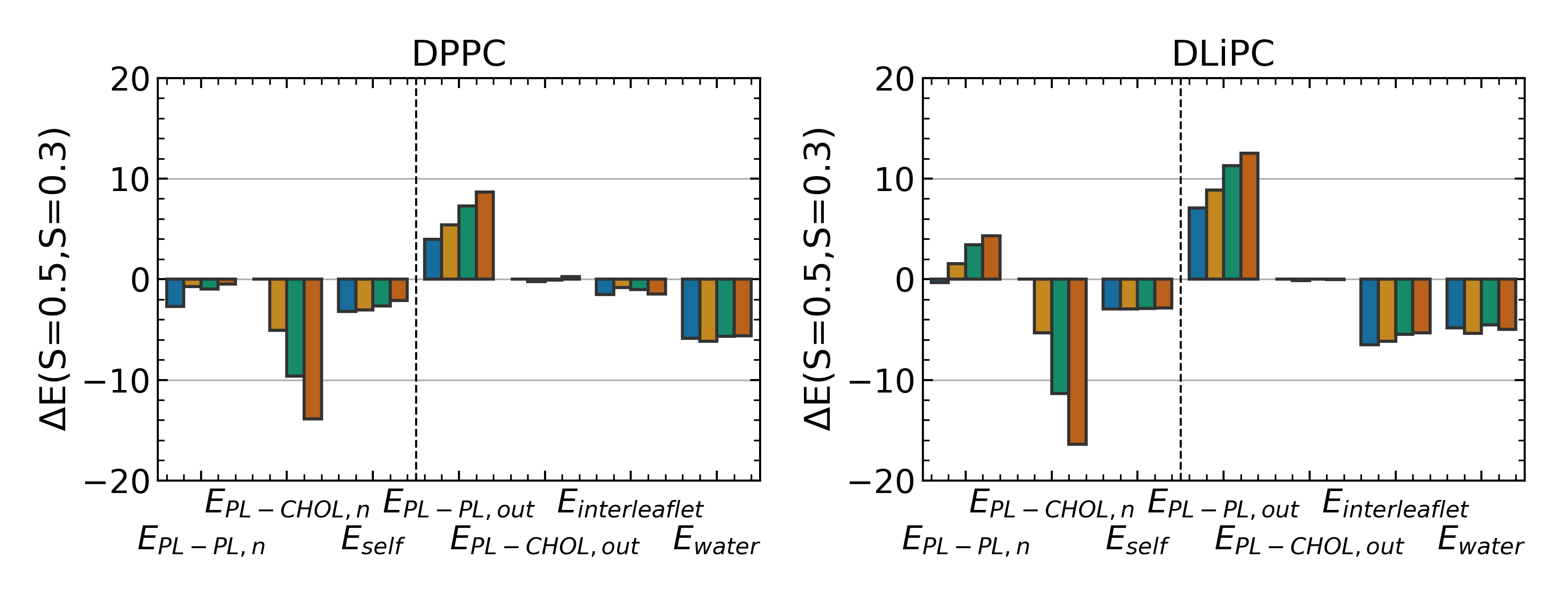}
	\caption{Contributions of the total interaction enthalpy gain for ordering the respective acyl chains from S=0.3 to S=0.5. The dashed line separates contributions from the nearest neighborhood of a lipid (left), from the non-nearest-neighbor contributions (right). Blue, yellow, green and red indicate that contributions were derived from bilayers with CHOL concentrations of 0, 10, 20 and 30 mol\% respectively.}
	\label{fig:Ecomp}
\end{figure*}

In figure~\ref{fig:Ecomp} we summarize the results obtained so far by comparing the change of enthalpies when increasing the order parameter in the relevant range from S=0.3 to S=0.5. This representation is sensitive to study the impact of CHOL. In this representation we acknowledge that the change in energy rather than the absolute value matters for the estimation of the resulting PL order parameters. We can distinguish three types of interaction terms with respect to the dependence on the CHOL concentration: (i) As expected $E_{PL-CHOL}$ is proportional to the CHOL concentration. (ii) $E_{PL-PL}$ displays a more complex dependence on the CHOL concentration where the strongest effects are present for $E_{PL-PL,out}$. (iii) The remaining terms are basically independent from the CHOL concentration. This distinction holds for both PL.

\begin{figure*}
    \centering
    \includegraphics[width=16cm]{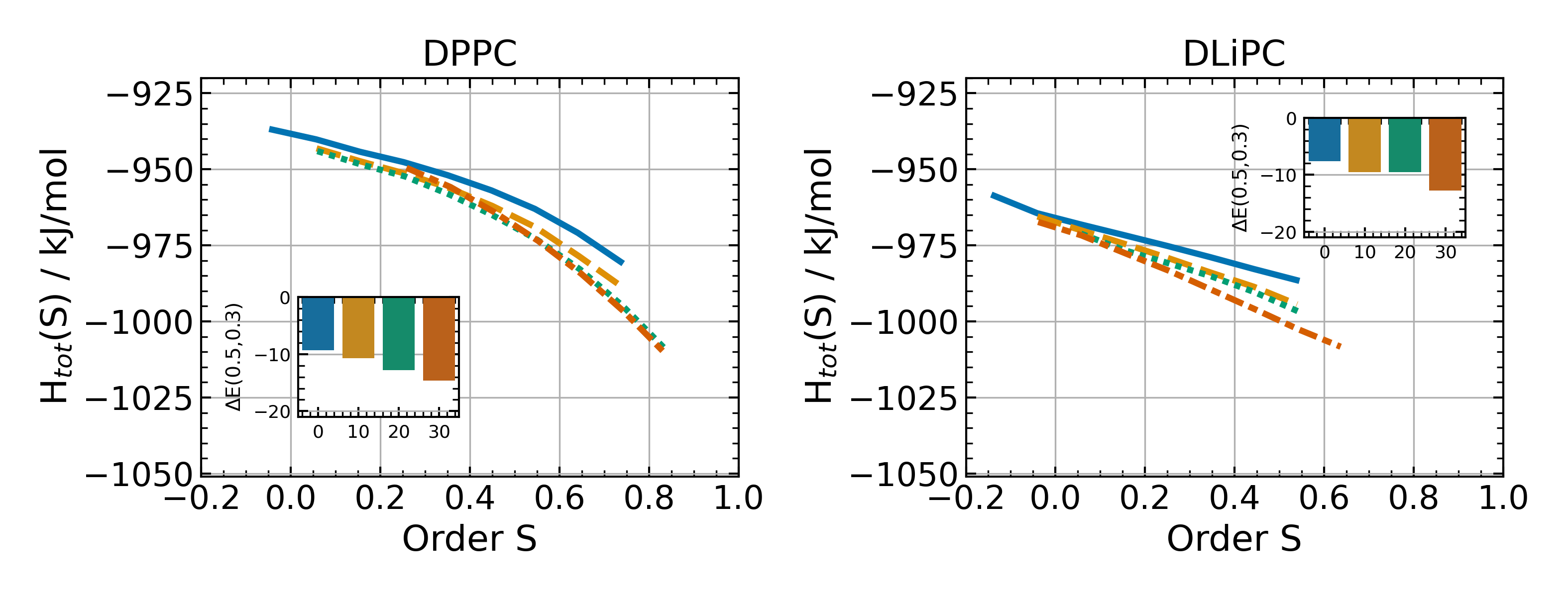}
    \caption{Sum of all interaction components $E_{PL-PL}$, $E_{PL-CHOL}$, $E_{interleaflet}$, $E_{water}$ and $E_{self}$ as a function of PL chain order parameter and CHOL concentration. The insets show the respective energy gain by changing the PL's chain order parameter from a disordered (S=0.3) to a more ordered state (S=0.5). Solid blue, dashed yellow, dotted green and dot-dashed red indicate that energies were derived from bilayers with CHOL concentrations of 0, 10, 20 and 30 mol\% respectively.}
    \label{fig:Etot}
\end{figure*}

Now we are in the position to estimate the overall enthalpic energy contribution for a randomly chosen PL by summing up all contributions, discussed so far. The result for the total energy $H_{tot}(S)$ is shown in figure \ref{fig:Etot}. Remarkably, the results for DPPC and DLiPC hardly differ in the order parameter range between 0.3 and 0.5. This statement holds for the dependence on the order parameter as well as on the CHOL concentration.  Note that individual energy contributions so far showed much larger differences between the saturated and the unsaturated lipids. For example, the interleaflet interaction showed very different dependencies on the order parameter for both lipids, whereas, e.g., the PL-CHOL interaction displayed a much larger dependence on CHOL concentration. Thus, we may conclude that there is a significant canceling effect among the different contributions.

\subsection{Chain entropy}

In figure \ref{fig:pscd} we showed that CHOL has an ordering effect on DPPC roughly twice as strong as on DLiPC, with DPPC exhibiting overall higher order parameter values than DLiPC. Indeed, the energy gain for a transition from a disordered to a (more) ordered state (S=0.3 to S=0.5) increases linearly with CHOL concentration. However, the energy gain for the two PL types is very similar so that the impact of CHOL on PL order cannot solely be understood via its effect on the enthalpy gain.

\begin{figure*}
    \centering
    \includegraphics[width=16cm]{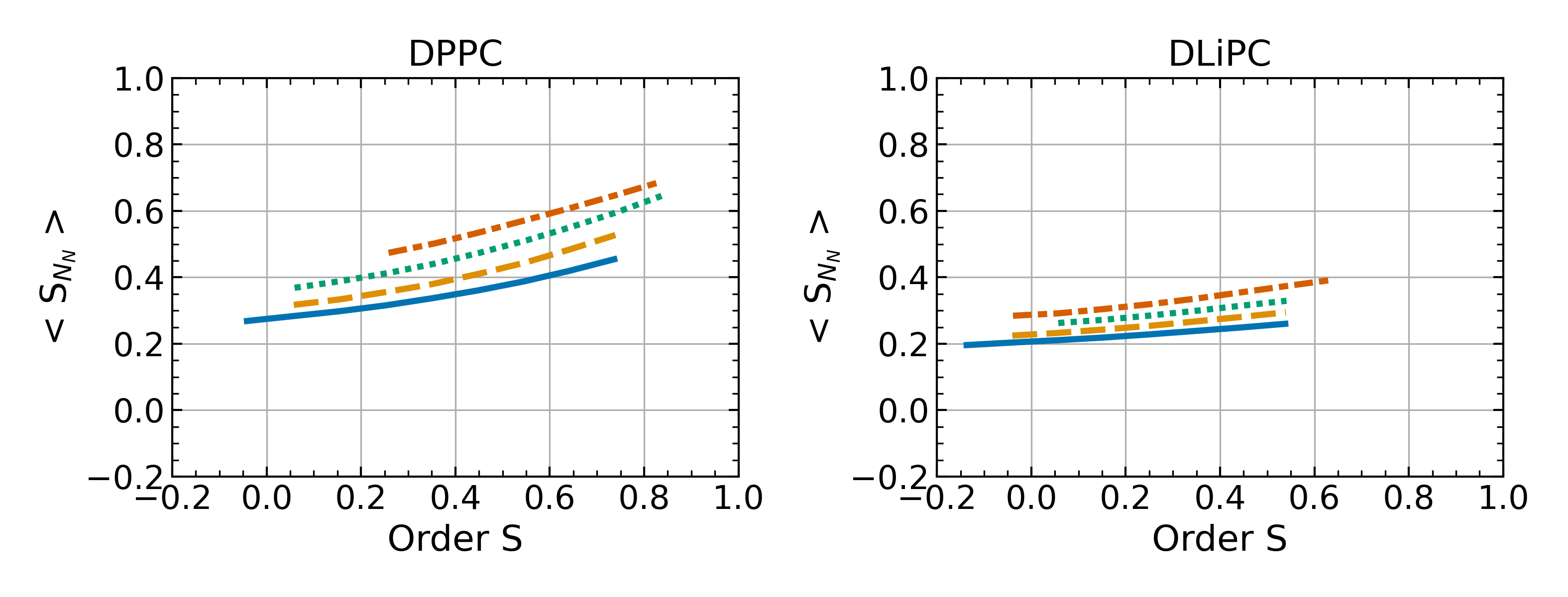}
    \caption{Average order parameter of the PL neighborhood as a function of the PL's order parameter. Solid blue, dashed yellow, dotted green and dot-dashed red indicate that energies were derived from bilayers with CHOL concentrations of 0, 10, 20 and 30 mol\% respectively.}
    \label{fig:fluctuation}
\end{figure*}

For a complete description of the system also chain entropy effects have to be taken into account. Qualitatively, they express that for entropic reasons the emergence of states with high order parameters is suppressed. Formally, the entropic effects can be captured by a function $Z(S)$, defined via the proportionality
\begin{equation}
p(S) \propto \exp(-\beta H_{tot}(S)) Z(S)
\label{eq:Boltzmann}
\end{equation}
where $\beta$ is the inverse product of the molar gas constant R and the temperature T.

A priori it is not clear that $Z(S)$ can be interpreted as the exponential of the entropy. Only if this interpretation is valid, knowledge of $Z(S)$ could be used for the prediction of, e.g., the temperature dependence of $p(S)$. There exist two  limits for which this interpretation is possible. They can be defined based on the relation between the order parameter $S_c$ of the central molecule, the average order parameter of the nearest neighbors $\langle S_{N_N} \rangle$ and the average order parameter $\langle S \rangle$ of the total system. For this purpose we define a constant $f_c$ via the relation
\begin{equation}
\langle S_{N_N} \rangle  - \langle S \rangle = f_c (S_c - \langle S \rangle).
\end{equation}

Now two limiting cases can be defined. Scenario 1: $f_S = 0$. This corresponds to the mean-field limit, yielding a simple one-particle picture in an average environment. Typically it is observed for (model) systems in high dimensions. Scenario 2: $f_S = 1$.  In this scenario the order parameter of the central particle characterizes the order parameter of the total system. Here it is possible to calculate the total energy of the system which can be written as $N \tilde{H}_{tot}(S)$. Note that $\tilde{H}_{tot}(S) \ne H_{tot}(S)$ because all PL-PL pair energies have to be reduced by a factor of 2 to avoid double counting ($N$: number of PL molecules).  If neither (1) or (2) are fulfilled, the overall enthalpy $H_{tot}(S)$ is difficult to be interpreted in a thermodynamic sense.

For our numerical data the function $\langle S_{N_N} \rangle$ is shown as a function of the order parameter of the central molecule in figure \ref{fig:fluctuation}. In the relevant order parameter range between $S=0.2$ and $S=0.5$ we find for DPPC $f_S = 0.2 - 0.3$ (depending on the CHOL concentration) and for DLiPC that $f_S =0.08-0.12$.  We may conclude that for DLiPC the system behaves close to the mean-field limit (albeit not exactly) whereas for DPPC much stronger correlation effects are present which render a thermodynamic interpretation, based on the total enthalpy, difficult.  Thus in the following we restrict ourselves to the discussion of DLiPC. For reasons of comparison the results for DPPC are shown in the SI.

\begin{figure}
    \centering
    \includegraphics[width=8cm]{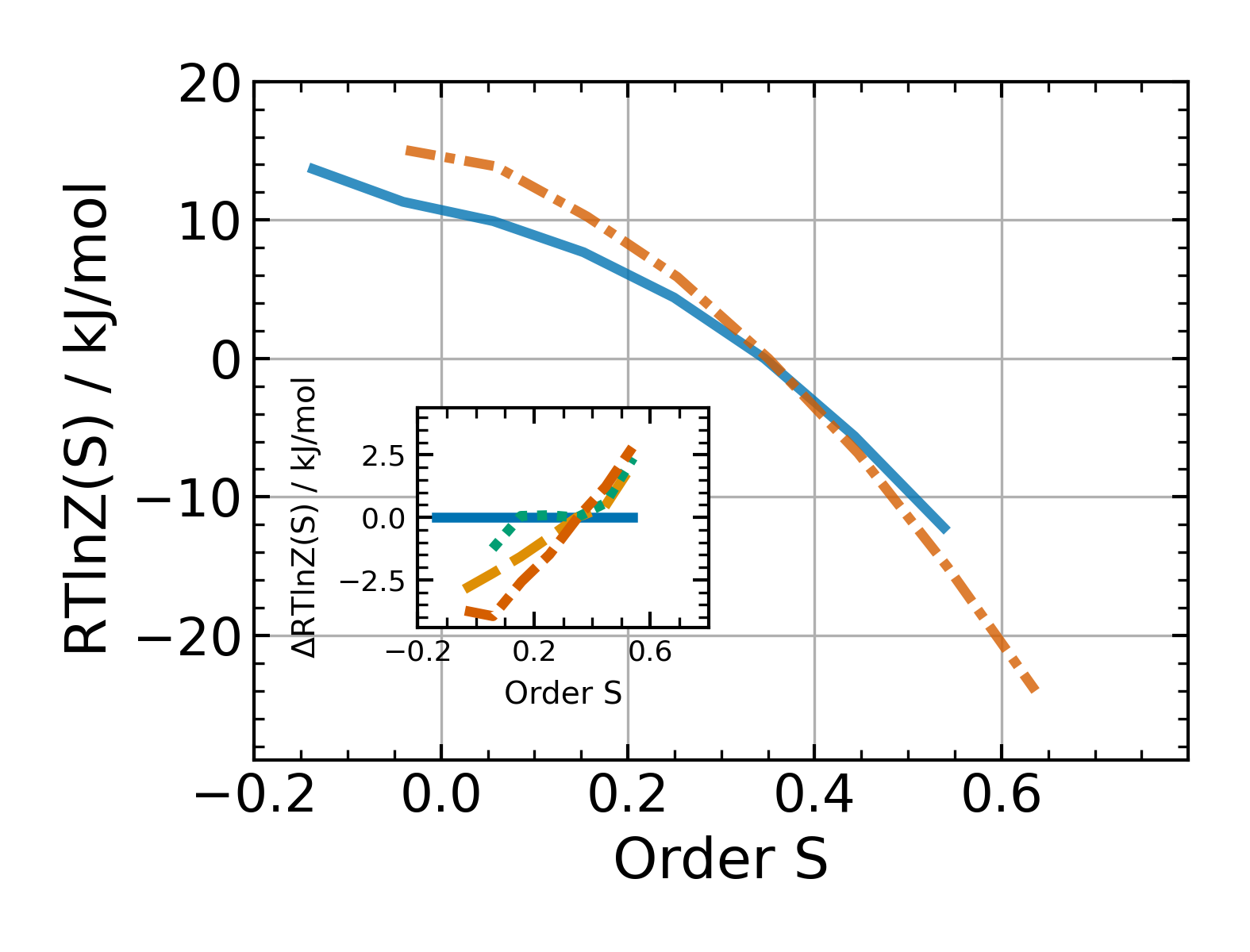}
    \caption{Estimation of the acyl chain entropy of DLiPC derived from the total sum of all enthalpic energy contributions and the order parameter distributions in the simulations at two CHOL concentrations and the difference between between the respective estimates in bilayer composition from 0\% to 30\% CHOL (inset). Solid blue, dashed yellow, dotted green and dot-dashed red indicate that energies were derived from bilayers with CHOL concentrations of 0, 10, 20 and 30 mol\% respectively.}
    \label{fig:zofs_DLiPC}
\end{figure}

The function $Z(S)$, obtained via $Z(S) \propto \exp(\beta H_{tot}(S)) p(S)$, is shown in figure~\ref{fig:zofs_DLiPC}. Similar to the interaction enthalpy functions, $Z(S)$ were averaged over temperatures of 290~K to 330~K for DLiPC, 330~K to 350~K for DPPC and, for better comparability, were shifted to zero at S=0.35, which roughly lies in the order parameter region of the pure PL bilayers in the disordered phase (DPPC S=0.35, DLiPC S=0.25) (figure~\ref{fig:zofs_DLiPC}, DLiPC and figure~\ref{fig:zofs_DPPC} in the SI).

\begin{figure}
    \centering
    \includegraphics[width=8cm]{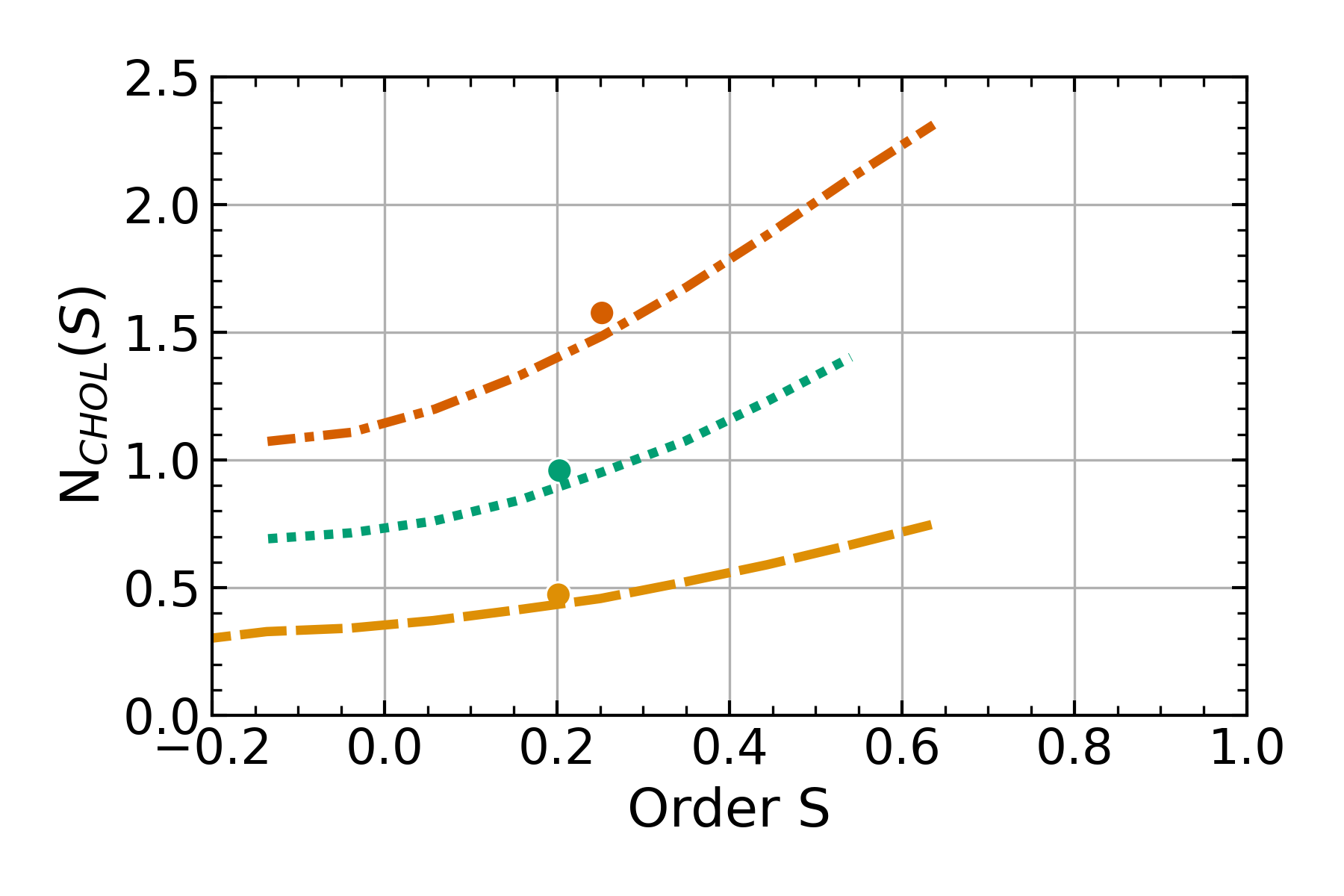}
    \caption{Number of nearest CHOL neighbors of DLiPC as a function of the its order parameter. The dots indicate the overall average number of nearest CHOL neighbors and order parameter.}
    \label{fig:NcofStogether}
\end{figure}

Naturally, one observes that high order parameters are strongly disfavored and, comparing the curves for the systems with and without CHOL, presence of CHOL increases the entropic difference between the ordered and disordered state. This dependence on CHOL concentration is further quantified via the comparison of $\Delta RT \ln Z(S)$ with the function at zero CHOL concentration (inset of figure \ref{fig:zofs_DLiPC}), using $T = 290$~K. Despite some fluctuations the same effect is also observed for the other CHOL concentrations. Generally, one may  expect  that the neighboorhood of a CHOL molecule reduces the number of available configurations for adjacent PL molecules for all order parameters. A priori it is not evident for which order parameters the reduction is higher or lower. On the one hand, one may expect that very disordered chains (low S) feel a strong reduction due to inaccessibility of the volume, where the rigid body of CHOL is located. On the other hand, for high order parameters there is a significant enthalpic interaction of the PL to the CHOL (see figure~\ref{fig:Eplc}), correspondingly reducing the conformational degrees of freeedom. However, there is another strong effect which can be derived from analyzing the number of CHOL neighbors around a PL as function of the order parameter; see figure~\ref{fig:NcofStogether}.  It turns out that CHOL is more likely located around PLs with high order parameters. Indeed, the proximity of CHOL may be one reason for a PL to display a higher order parameter. This observation naturally explains the suppression of entropy for high order parameters upon adding CHOL, seen in figure~\ref{fig:zofs_DLiPC}. Finally, we mention that the overall dependence on the CHOL concentration is of the order of 5 kJ/mol. This is close to the CHOL dependence of $H_{tot}(S)$, as shown in figure~\ref{fig:Etot}. Thus, the entropic and enthalpic contribution are of similar importance when discussing the impact of CHOL.

We refrain from a closer discussion of the function $Z_{DPPC}(S)$ due to reasons mentioned above. We just mention that for DPPC the dependence on the order parameter is somewhat weaker and that the dependence on the CHOL concentration is basically absent; see SI.

\begin{figure}
    \centering
    \includegraphics[width=8cm]{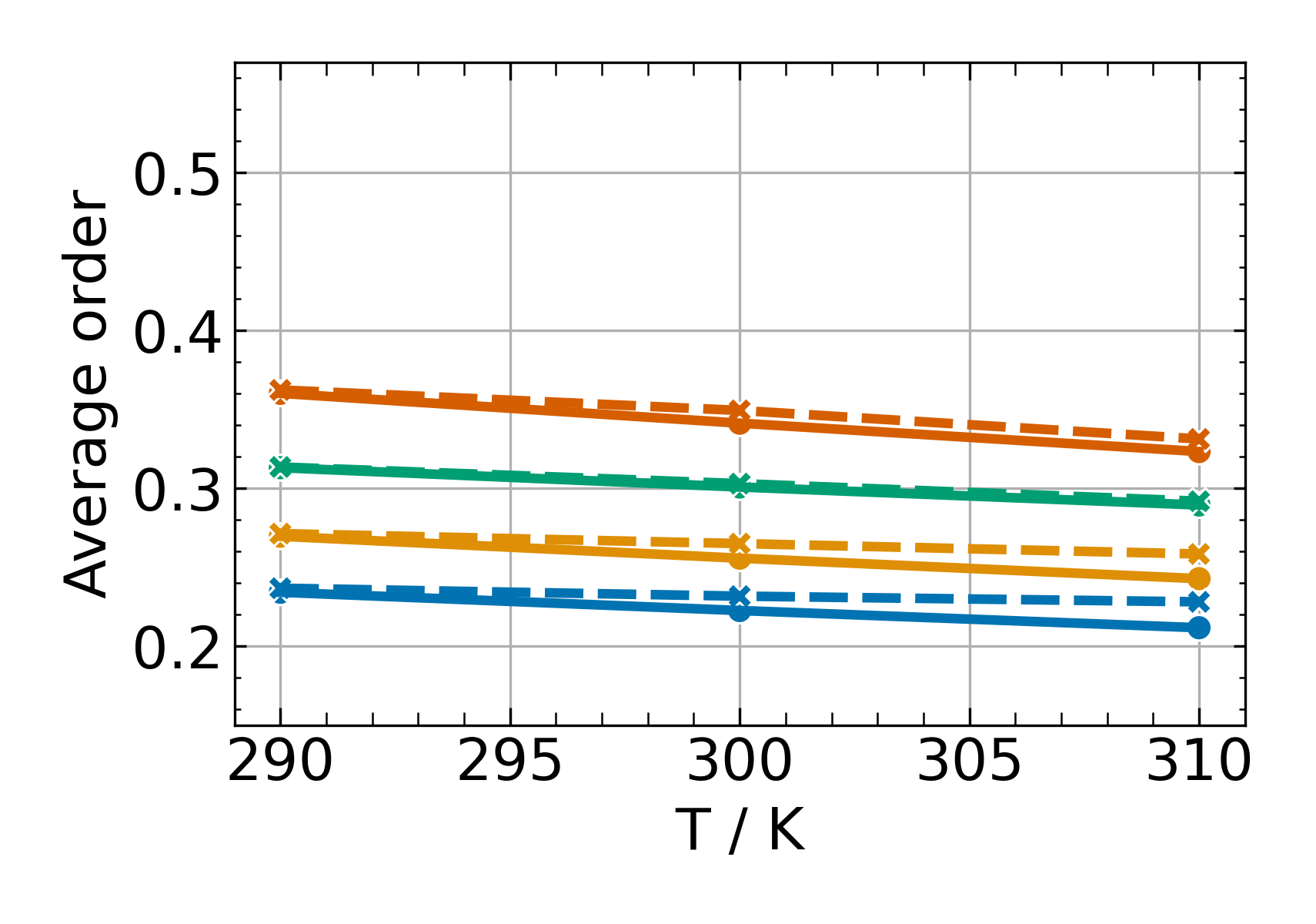}
    \caption{Average order parameter as a function of simulation temperature (dashed) at concentrations of 0, 10, 20 and 30 mol\% CHOL (blue, yellow, green and red respectively) and the recalculated average order parameter from Z(S) derived at from the simulation at 290~K (solid).}
    \label{fig:pfromz_DLiPC_Tdep}
\end{figure}

Having found a decomposition in the enthalpic and the entropic contributions for DLiPC as a function of order parameter, it is possible to estimate properties of the system as a function of temperature.  More specifically, we estimated $Z(S)$ via Eq.\ref{eq:Boltzmann} for a temperature of 290~K and predicted the average order parameter as a function of temperature. The results for the different CHOL concentrations are shown in figure~\ref{fig:pfromz_DLiPC_Tdep}. Indeed, the temperature dependence is reproduced very well for higher CHOL concentrations. At lower CHOL concentrations deviations between prediction and simulation are observed. Of course, deviations can be naturally explained because, first, the mean-field approximation does not perfectly hold and, second, in general the entropy is temperature dependent.

Finally, one can wonder whether the difference of $H_{tot}(S)$ between DPPC and DLiPC is relevant for the thermodynamic properties. For this purpose we have calculated the DLiPC average order parameter as a function of CHOL concentration also when deliberately using $H_{tot}(S)$, obtained for DPPC, for predicting the DLiPC properties. As shown in the SI there are indeed dramatic differences in particular for high CHOL concentrations (prediction of an average order parameter of 0.5 as compared to the correct value of 0.3). Thus, the difference of $H_{tot}(S)$ between the saturated and unsaturated PL, although quite small from visual inspection, is still significant.

\section{Conclusion}

In this work, we have derived the different contributions to the total interaction energy of cholesterol (CHOL) ordering of DPPC or DLiPC to understand the differing affinity exhibited by CHOL, as well as the underlying mechanism of cholesterol's unique ordering capability. Having studied the prototype lipids DPPC and DLiPC as proxies for saturated and unsaturated lipids, we believe that our results provide important new insight for the interplay of the different enthalpy contributions of membranes containing CHOL and phospholipids.

A key observation is the relative strength of the PL-CHOL interaction with respect to the PL-PL interaction: The energetic contribution from the PL-CHOL interaction for ordering DPPC or DLiPC from S=0.3 to S=0.5 is roughly 2 to 3 times larger. Additionally, the maximum in interaction strength of the PL-CHOL interaction as a function of the order parameter is a cornerstone for understanding the simultaneous effect of CHOL to increase the order in the disordered phase and to fluidify the gel state. Furthermore, we could rationalize why inherently disordered lipids display a stronger interaction between the bilayer leaflets and show that CHOL only weakly influences the bilayer-water interaction, independent of the lipid type.

Finally, we derived an entropic contribution from the enthalpic interaction functions for DLiPC. With these entropic interaction functions along with the enthalpic contributions we were able to properly reproduce the temperature dependence. Additionally, with increasing CHOL concentration we found the entropic penalty of ordering DLiPC chains to increase significantly. We explain this behavior by CHOL's affinity for acyl chains with high order, thus mainly influencing this order regime.

Naturally, a complete thermodynamic understanding of the PL-CHOL mixtures requires the consideration of all pair-wise interactions so that the entropy-related function $Z(S)$ can be determined beyond the two limits discussed in this work. First steps in this direction, albeit without sterols, can be already found in literature.\cite{Hakobyan2017,Hakobyan2019}



\section*{Conflicts of interest}
There are no conflicts to declare.

\section*{Acknowledgement}
The authors thank Roland Wedlich-Söldner for helpful discussions and the Deutsche Forschungsgemeinschaft (DFG) for funding via SFB 1348.

\renewcommand\refname{References}

\bibliographystyle{rsc} 
\bibliography{./f_kell07_rcs}

\providecommand*{\mcitethebibliography}{\thebibliography}
\csname @ifundefined\endcsname{endmcitethebibliography}
{\let\endmcitethebibliography\endthebibliography}{}
\begin{mcitethebibliography}{42}
\providecommand*{\natexlab}[1]{#1}
\providecommand*{\mciteSetBstSublistMode}[1]{}
\providecommand*{\mciteSetBstMaxWidthForm}[2]{}
\providecommand*{\mciteBstWouldAddEndPuncttrue}
  {\def\EndOfBibitem{\unskip.}}
\providecommand*{\mciteBstWouldAddEndPunctfalse}
  {\let\EndOfBibitem\relax}
\providecommand*{\mciteSetBstMidEndSepPunct}[3]{}
\providecommand*{\mciteSetBstSublistLabelBeginEnd}[3]{}
\providecommand*{\EndOfBibitem}{}
\mciteSetBstSublistMode{f}
\mciteSetBstMaxWidthForm{subitem}
{(\emph{\alph{mcitesubitemcount}})}
\mciteSetBstSublistLabelBeginEnd{\mcitemaxwidthsubitemform\space}
{\relax}{\relax}

\bibitem[Sackmann(1995)]{Sackmann1995}
E.~Sackmann, in \emph{Structure and Dynamics of Membranes}, ed. R.~Lipowsky and
  E.~Sackmann, North-Holland, 1995, vol.~1 of Handbook of Biological Physics,
  pp. 1--63\relax
\mciteBstWouldAddEndPuncttrue
\mciteSetBstMidEndSepPunct{\mcitedefaultmidpunct}
{\mcitedefaultendpunct}{\mcitedefaultseppunct}\relax
\EndOfBibitem
\bibitem[Sezgin \emph{et~al.}(2017)Sezgin, Levental, Mayor, and
  Eggeling]{Sezgin2017}
E.~Sezgin, I.~Levental, S.~Mayor and C.~Eggeling, \emph{Nat. Rev. Mol. Cell
  Biol.}, 2017, \textbf{18}, 361--374\relax
\mciteBstWouldAddEndPuncttrue
\mciteSetBstMidEndSepPunct{\mcitedefaultmidpunct}
{\mcitedefaultendpunct}{\mcitedefaultseppunct}\relax
\EndOfBibitem
\bibitem[Levental and Veatch(2016)]{Levental2016}
I.~Levental and S.~L. Veatch, \emph{J. Mol. Biol.}, 2016, \textbf{428},
  4749--4764\relax
\mciteBstWouldAddEndPuncttrue
\mciteSetBstMidEndSepPunct{\mcitedefaultmidpunct}
{\mcitedefaultendpunct}{\mcitedefaultseppunct}\relax
\EndOfBibitem
\bibitem[Róg and Vattulainen(2014)]{Rog2014}
T.~Róg and I.~Vattulainen, \emph{Chem. Phys. Lipids}, 2014, \textbf{184},
  82--104\relax
\mciteBstWouldAddEndPuncttrue
\mciteSetBstMidEndSepPunct{\mcitedefaultmidpunct}
{\mcitedefaultendpunct}{\mcitedefaultseppunct}\relax
\EndOfBibitem
\bibitem[Simons and Sampaio(2011)]{Simons2011}
K.~Simons and J.~L. Sampaio, \emph{Cold Spring Harbor Perspect. Biol.}, 2011,
  \textbf{3}, 1--17\relax
\mciteBstWouldAddEndPuncttrue
\mciteSetBstMidEndSepPunct{\mcitedefaultmidpunct}
{\mcitedefaultendpunct}{\mcitedefaultseppunct}\relax
\EndOfBibitem
\bibitem[Heberle and Feigenson(2011)]{Heberle2011}
F.~A. Heberle and G.~W. Feigenson, \emph{Cold Spring Harbor Perspect. Biol.},
  2011, \textbf{3}, 1--13\relax
\mciteBstWouldAddEndPuncttrue
\mciteSetBstMidEndSepPunct{\mcitedefaultmidpunct}
{\mcitedefaultendpunct}{\mcitedefaultseppunct}\relax
\EndOfBibitem
\bibitem[Fan \emph{et~al.}(2010)Fan, Sammalkorpi, and Haataja]{Fan2010}
J.~Fan, M.~Sammalkorpi and M.~Haataja, \emph{FEBS Lett.}, 2010, \textbf{584},
  1678--1684\relax
\mciteBstWouldAddEndPuncttrue
\mciteSetBstMidEndSepPunct{\mcitedefaultmidpunct}
{\mcitedefaultendpunct}{\mcitedefaultseppunct}\relax
\EndOfBibitem
\bibitem[Róg \emph{et~al.}(2009)Róg, Pasenkiewicz-Gierula, Vattulainen, and
  Karttunen]{Rog2009}
T.~Róg, M.~Pasenkiewicz-Gierula, I.~Vattulainen and M.~Karttunen,
  \emph{Biochim. Biophys. Acta, Biomembr.}, 2009, \textbf{1788}, 97--121\relax
\mciteBstWouldAddEndPuncttrue
\mciteSetBstMidEndSepPunct{\mcitedefaultmidpunct}
{\mcitedefaultendpunct}{\mcitedefaultseppunct}\relax
\EndOfBibitem
\bibitem[van Meer \emph{et~al.}(2008)van Meer, Voelker, and
  Feigenson]{Meer2008}
G.~van Meer, D.~R. Voelker and G.~W. Feigenson, \emph{Nat. Rev. Mol. Cell
  Biol.}, 2008, \textbf{9}, 112--124\relax
\mciteBstWouldAddEndPuncttrue
\mciteSetBstMidEndSepPunct{\mcitedefaultmidpunct}
{\mcitedefaultendpunct}{\mcitedefaultseppunct}\relax
\EndOfBibitem
\bibitem[Pan \emph{et~al.}(2008)Pan, Mills, Tristram-Nagle, and Nagle]{Pan2008}
J.~Pan, T.~T. Mills, S.~Tristram-Nagle and J.~F. Nagle, \emph{Phys. Rev.
  Lett.}, 2008, \textbf{100}, 198103\relax
\mciteBstWouldAddEndPuncttrue
\mciteSetBstMidEndSepPunct{\mcitedefaultmidpunct}
{\mcitedefaultendpunct}{\mcitedefaultseppunct}\relax
\EndOfBibitem
\bibitem[Pan \emph{et~al.}(2009)Pan, Tristram-Nagle, and Nagle]{Pan2009}
J.~Pan, S.~Tristram-Nagle and J.~F. Nagle, \emph{Phys. Rev. E}, 2009,
  \textbf{80}, 021931\relax
\mciteBstWouldAddEndPuncttrue
\mciteSetBstMidEndSepPunct{\mcitedefaultmidpunct}
{\mcitedefaultendpunct}{\mcitedefaultseppunct}\relax
\EndOfBibitem
\bibitem[Engberg \emph{et~al.}(2016)Engberg, Hautala, Yasuda, Dehio, Murata,
  Slotte, and Nyholm]{Engberg2016}
O.~Engberg, V.~Hautala, T.~Yasuda, H.~Dehio, M.~Murata, J.~P. Slotte and T.~K.
  Nyholm, \emph{Biophys. J.}, 2016, \textbf{111}, 546--556\relax
\mciteBstWouldAddEndPuncttrue
\mciteSetBstMidEndSepPunct{\mcitedefaultmidpunct}
{\mcitedefaultendpunct}{\mcitedefaultseppunct}\relax
\EndOfBibitem
\bibitem[Yang \emph{et~al.}(2016)Yang, Martí, and Calero]{Yang2016}
J.~Yang, J.~Martí and C.~Calero, \emph{Soft Matter}, 2016, \textbf{12},
  4557--4561\relax
\mciteBstWouldAddEndPuncttrue
\mciteSetBstMidEndSepPunct{\mcitedefaultmidpunct}
{\mcitedefaultendpunct}{\mcitedefaultseppunct}\relax
\EndOfBibitem
\bibitem[Chakraborty \emph{et~al.}(2020)Chakraborty, Doktorova, Molugu,
  Heberle, Scott, Dzikovski, Nagao, Stingaciu, Standaert, Barrera, Katsaras,
  Khelashvili, Brown, and Ashkar]{Chakraborty2020}
S.~Chakraborty, M.~Doktorova, T.~R. Molugu, F.~A. Heberle, H.~L. Scott,
  B.~Dzikovski, M.~Nagao, L.~R. Stingaciu, R.~F. Standaert, F.~N. Barrera,
  J.~Katsaras, G.~Khelashvili, M.~F. Brown and R.~Ashkar, \emph{Proc. Natl.
  Acad. Sci. U. S. A.}, 2020, \textbf{117}, 21896--21905\relax
\mciteBstWouldAddEndPuncttrue
\mciteSetBstMidEndSepPunct{\mcitedefaultmidpunct}
{\mcitedefaultendpunct}{\mcitedefaultseppunct}\relax
\EndOfBibitem
\bibitem[Huang and Feigenson(1999)]{Huang1999}
J.~Huang and G.~W. Feigenson, \emph{Biophys. J.}, 1999, \textbf{76},
  2142--2157\relax
\mciteBstWouldAddEndPuncttrue
\mciteSetBstMidEndSepPunct{\mcitedefaultmidpunct}
{\mcitedefaultendpunct}{\mcitedefaultseppunct}\relax
\EndOfBibitem
\bibitem[Daly \emph{et~al.}(2011)Daly, Wang, and Regen]{Daly2011}
T.~A. Daly, M.~Wang and S.~L. Regen, \emph{Langmuir}, 2011, \textbf{27},
  2159--2161\relax
\mciteBstWouldAddEndPuncttrue
\mciteSetBstMidEndSepPunct{\mcitedefaultmidpunct}
{\mcitedefaultendpunct}{\mcitedefaultseppunct}\relax
\EndOfBibitem
\bibitem[Krause and Regen(2014)]{Krause2014}
M.~R. Krause and S.~L. Regen, \emph{Acc. Chem. Res.}, 2014, \textbf{47},
  3512--3521\relax
\mciteBstWouldAddEndPuncttrue
\mciteSetBstMidEndSepPunct{\mcitedefaultmidpunct}
{\mcitedefaultendpunct}{\mcitedefaultseppunct}\relax
\EndOfBibitem
\bibitem[Mukai and Regen(2017)]{Mukai2017}
M.~Mukai and S.~L. Regen, \emph{Bull. Chem. Soc. Jpn.}, 2017, \textbf{90},
  1083--1087\relax
\mciteBstWouldAddEndPuncttrue
\mciteSetBstMidEndSepPunct{\mcitedefaultmidpunct}
{\mcitedefaultendpunct}{\mcitedefaultseppunct}\relax
\EndOfBibitem
\bibitem[Wang \emph{et~al.}(2018)Wang, Almeida, and Regen]{Wang2018}
C.~Wang, P.~F. Almeida and S.~L. Regen, \emph{Biochemistry}, 2018, \textbf{57},
  6637--6643\relax
\mciteBstWouldAddEndPuncttrue
\mciteSetBstMidEndSepPunct{\mcitedefaultmidpunct}
{\mcitedefaultendpunct}{\mcitedefaultseppunct}\relax
\EndOfBibitem
\bibitem[Polley \emph{et~al.}(2014)Polley, Mayor, and Rao]{Polley2014}
A.~Polley, S.~Mayor and M.~Rao, \emph{J. Chem. Phys.}, 2014, \textbf{141},
  064903\relax
\mciteBstWouldAddEndPuncttrue
\mciteSetBstMidEndSepPunct{\mcitedefaultmidpunct}
{\mcitedefaultendpunct}{\mcitedefaultseppunct}\relax
\EndOfBibitem
\bibitem[Seo \emph{et~al.}(2020)Seo, Murata, and Shinoda]{Seo2020}
S.~Seo, M.~Murata and W.~Shinoda, \emph{J. Phys. Chem. Lett.}, 2020,
  \textbf{11}, 5171--5176\relax
\mciteBstWouldAddEndPuncttrue
\mciteSetBstMidEndSepPunct{\mcitedefaultmidpunct}
{\mcitedefaultendpunct}{\mcitedefaultseppunct}\relax
\EndOfBibitem
\bibitem[Tian \emph{et~al.}(2016)Tian, Nickels, Katsaras, and Cheng]{Tian2016}
J.~Tian, J.~Nickels, J.~Katsaras and X.~Cheng, \emph{J. Phys. Chem. B}, 2016,
  \textbf{120}, 8438--8448\relax
\mciteBstWouldAddEndPuncttrue
\mciteSetBstMidEndSepPunct{\mcitedefaultmidpunct}
{\mcitedefaultendpunct}{\mcitedefaultseppunct}\relax
\EndOfBibitem
\bibitem[Zhang and Lin(2019)]{Zhang2019}
S.~Zhang and X.~Lin, \emph{J. Am. Chem. Soc.}, 2019, \textbf{141},
  15884--15890\relax
\mciteBstWouldAddEndPuncttrue
\mciteSetBstMidEndSepPunct{\mcitedefaultmidpunct}
{\mcitedefaultendpunct}{\mcitedefaultseppunct}\relax
\EndOfBibitem
\bibitem[Leeb and Maibaum(2018)]{Leeb2018}
F.~Leeb and L.~Maibaum, \emph{Biophys. J.}, 2018, \textbf{115},
  2179--2188\relax
\mciteBstWouldAddEndPuncttrue
\mciteSetBstMidEndSepPunct{\mcitedefaultmidpunct}
{\mcitedefaultendpunct}{\mcitedefaultseppunct}\relax
\EndOfBibitem
\bibitem[McIntosh(1978)]{McIntosh1978}
T.~J. McIntosh, \emph{Biochim. Biophys. Acta, Biomembr.}, 1978, \textbf{513},
  43--58\relax
\mciteBstWouldAddEndPuncttrue
\mciteSetBstMidEndSepPunct{\mcitedefaultmidpunct}
{\mcitedefaultendpunct}{\mcitedefaultseppunct}\relax
\EndOfBibitem
\bibitem[Subczynski \emph{et~al.}(1994)Subczynski, Wisniewska, Subczynski,
  Wisniewska, Yin, Hyde, and Kusumi]{Subczynski1994}
W.~K. Subczynski, A.~Wisniewska, W.~K. Subczynski, A.~Wisniewska, J.~J. Yin,
  J.~S. Hyde and A.~Kusumi, \emph{Biochemistry}, 1994, \textbf{33},
  7670--7681\relax
\mciteBstWouldAddEndPuncttrue
\mciteSetBstMidEndSepPunct{\mcitedefaultmidpunct}
{\mcitedefaultendpunct}{\mcitedefaultseppunct}\relax
\EndOfBibitem
\bibitem[Pasenkiewicz-Gierula \emph{et~al.}(2000)Pasenkiewicz-Gierula, Róg,
  Kitamura, and Kusumi]{Pasenkiewicz2000}
M.~Pasenkiewicz-Gierula, T.~Róg, K.~Kitamura and A.~Kusumi, \emph{Biophys.
  J.}, 2000, \textbf{78}, 1376--1389\relax
\mciteBstWouldAddEndPuncttrue
\mciteSetBstMidEndSepPunct{\mcitedefaultmidpunct}
{\mcitedefaultendpunct}{\mcitedefaultseppunct}\relax
\EndOfBibitem
\bibitem[Pluhackova \emph{et~al.}(2016)Pluhackova, Kirsch, Han, Sun, Jiang,
  Unruh, and Böckmann]{Pluhackova2016}
K.~Pluhackova, S.~A. Kirsch, J.~Han, L.~Sun, Z.~Jiang, T.~Unruh and R.~A.
  Böckmann, \emph{J. Phys. Chem. B}, 2016, \textbf{120}, 3888--3903\relax
\mciteBstWouldAddEndPuncttrue
\mciteSetBstMidEndSepPunct{\mcitedefaultmidpunct}
{\mcitedefaultendpunct}{\mcitedefaultseppunct}\relax
\EndOfBibitem
\bibitem[Favela-Rosales \emph{et~al.}(2020)Favela-Rosales,
  Galv\'{a}n-Hern\'{a}ndez, Hern\'{a}ndez-Cobos, Kobayashi, Carbajal-Tinoco,
  Nakabayashi, and Ortega-Blake]{Rosales2020}
F.~Favela-Rosales, A.~Galv\'{a}n-Hern\'{a}ndez, J.~Hern\'{a}ndez-Cobos,
  N.~Kobayashi, M.~D. Carbajal-Tinoco, S.~Nakabayashi and I.~Ortega-Blake,
  \emph{Biophys. Chem.}, 2020, \textbf{257}, 106275\relax
\mciteBstWouldAddEndPuncttrue
\mciteSetBstMidEndSepPunct{\mcitedefaultmidpunct}
{\mcitedefaultendpunct}{\mcitedefaultseppunct}\relax
\EndOfBibitem
\bibitem[Lee \emph{et~al.}(2016)Lee, Cheng, Swails, Yeom, Eastman, Lemkul, Wei,
  Buckner, Jeong, Qi, Jo, Pande, Case, Brooks, MacKerell, Klauda, and
  Im]{Lee2016}
J.~Lee, X.~Cheng, J.~M. Swails, M.~S. Yeom, P.~K. Eastman, J.~A. Lemkul,
  S.~Wei, J.~Buckner, J.~C. Jeong, Y.~Qi, S.~Jo, V.~S. Pande, D.~A. Case, C.~L.
  Brooks, A.~D. MacKerell, J.~B. Klauda and W.~Im, \emph{J. Chem. Theory
  Comput.}, 2016, \textbf{12}, 405--413\relax
\mciteBstWouldAddEndPuncttrue
\mciteSetBstMidEndSepPunct{\mcitedefaultmidpunct}
{\mcitedefaultendpunct}{\mcitedefaultseppunct}\relax
\EndOfBibitem
\bibitem[Klauda \emph{et~al.}(2010)Klauda, Venable, Freites, O'Connor, Tobias,
  Mondragon-Ramirez, Vorobyov, MacKerell, and Pastor]{Klauda2010a}
J.~B. Klauda, R.~M. Venable, J.~A. Freites, J.~W. O'Connor, D.~J. Tobias,
  C.~Mondragon-Ramirez, I.~Vorobyov, A.~D. MacKerell and R.~W. Pastor, \emph{J.
  Phys. Chem. B}, 2010, \textbf{114}, 7830--7843\relax
\mciteBstWouldAddEndPuncttrue
\mciteSetBstMidEndSepPunct{\mcitedefaultmidpunct}
{\mcitedefaultendpunct}{\mcitedefaultseppunct}\relax
\EndOfBibitem
\bibitem[Abraham \emph{et~al.}(2015)Abraham, Murtola, Schulz, Páll, Smith,
  Hess, and Lindah]{Abraham2015}
M.~J. Abraham, T.~Murtola, R.~Schulz, S.~Páll, J.~C. Smith, B.~Hess and
  E.~Lindah, \emph{SoftwareX}, 2015, \textbf{1-2}, 19--25\relax
\mciteBstWouldAddEndPuncttrue
\mciteSetBstMidEndSepPunct{\mcitedefaultmidpunct}
{\mcitedefaultendpunct}{\mcitedefaultseppunct}\relax
\EndOfBibitem
\bibitem[Jorgensen \emph{et~al.}(1983)Jorgensen, Chandrasekhar, Madura, Impey,
  and Klein]{Jorgensen1983}
W.~L. Jorgensen, J.~Chandrasekhar, J.~D. Madura, R.~W. Impey and M.~L. Klein,
  \emph{J. Chem. Phys.}, 1983, \textbf{79}, 926--935\relax
\mciteBstWouldAddEndPuncttrue
\mciteSetBstMidEndSepPunct{\mcitedefaultmidpunct}
{\mcitedefaultendpunct}{\mcitedefaultseppunct}\relax
\EndOfBibitem
\bibitem[Nosé(1984)]{Nose1984}
S.~Nosé, \emph{J. Chem. Phys.}, 1984, \textbf{81}, 511--519\relax
\mciteBstWouldAddEndPuncttrue
\mciteSetBstMidEndSepPunct{\mcitedefaultmidpunct}
{\mcitedefaultendpunct}{\mcitedefaultseppunct}\relax
\EndOfBibitem
\bibitem[Parrinello and Rahman(1981)]{Parrinello1981}
M.~Parrinello and A.~Rahman, \emph{J. Appl. Phys. }, 1981, \textbf{52},
  7182--7190\relax
\mciteBstWouldAddEndPuncttrue
\mciteSetBstMidEndSepPunct{\mcitedefaultmidpunct}
{\mcitedefaultendpunct}{\mcitedefaultseppunct}\relax
\EndOfBibitem
\bibitem[Hess \emph{et~al.}(1997)Hess, Bekker, Berendsen, and
  Fraaije]{Hess1997}
B.~Hess, H.~Bekker, H.~J. Berendsen and J.~G. Fraaije, \emph{J. Comp. Chem.},
  1997, \textbf{18}, 1463--1472\relax
\mciteBstWouldAddEndPuncttrue
\mciteSetBstMidEndSepPunct{\mcitedefaultmidpunct}
{\mcitedefaultendpunct}{\mcitedefaultseppunct}\relax
\EndOfBibitem
\bibitem[Darden \emph{et~al.}(1993)Darden, York, and Pedersen]{Darden1993}
T.~Darden, D.~York and L.~Pedersen, \emph{J. Chem. Phys.}, 1993, \textbf{98},
  10089--10092\relax
\mciteBstWouldAddEndPuncttrue
\mciteSetBstMidEndSepPunct{\mcitedefaultmidpunct}
{\mcitedefaultendpunct}{\mcitedefaultseppunct}\relax
\EndOfBibitem
\bibitem[Michaud-Agrawal \emph{et~al.}(2011)Michaud-Agrawal, Denning, Woolf,
  and Beckstein]{Agrawal2011}
N.~Michaud-Agrawal, E.~J. Denning, T.~B. Woolf and O.~Beckstein, \emph{J. Comp.
  Chem.}, 2011, \textbf{32}, 2319--2327\relax
\mciteBstWouldAddEndPuncttrue
\mciteSetBstMidEndSepPunct{\mcitedefaultmidpunct}
{\mcitedefaultendpunct}{\mcitedefaultseppunct}\relax
\EndOfBibitem
\bibitem[{R}ichard {J}.~{G}owers \emph{et~al.}(2016){R}ichard {J}.~{G}owers,
  {M}ax {L}inke, {J}onathan {B}arnoud, {T}yler {J}. {E}.~{R}eddy, {M}anuel
  {N}.~{M}elo, {S}ean {L}.~{S}eyler, {J}an {D}omański, {D}avid {L}.~{D}otson,
  {S}ébastien {B}uchoux, {I}an {M}.~{K}enney, and {O}liver
  {B}eckstein]{Gowers2016}
{R}ichard {J}.~{G}owers, {M}ax {L}inke, {J}onathan {B}arnoud, {T}yler {J}.
  {E}.~{R}eddy, {M}anuel {N}.~{M}elo, {S}ean {L}.~{S}eyler, {J}an {D}omański,
  {D}avid {L}.~{D}otson, {S}ébastien {B}uchoux, {I}an {M}.~{K}enney and
  {O}liver {B}eckstein, {P}roceedings of the 15th {P}ython in {S}cience
  {C}onference, 2016, pp. 98 -- 105\relax
\mciteBstWouldAddEndPuncttrue
\mciteSetBstMidEndSepPunct{\mcitedefaultmidpunct}
{\mcitedefaultendpunct}{\mcitedefaultseppunct}\relax
\EndOfBibitem
\bibitem[Humphrey \emph{et~al.}(1996)Humphrey, Dalke, and
  Schulten]{Humphrey1996}
W.~Humphrey, A.~Dalke and K.~Schulten, \emph{J. Mol. Graphics}, 1996,
  \textbf{14}, 33--38\relax
\mciteBstWouldAddEndPuncttrue
\mciteSetBstMidEndSepPunct{\mcitedefaultmidpunct}
{\mcitedefaultendpunct}{\mcitedefaultseppunct}\relax
\EndOfBibitem
\bibitem[Hakobyan and Heuer(2017)]{Hakobyan2017}
D.~Hakobyan and A.~Heuer, \emph{J. Chem. Phys.}, 2017, \textbf{146},
  064305\relax
\mciteBstWouldAddEndPuncttrue
\mciteSetBstMidEndSepPunct{\mcitedefaultmidpunct}
{\mcitedefaultendpunct}{\mcitedefaultseppunct}\relax
\EndOfBibitem
\bibitem[Hakobyan and Heuer(2019)]{Hakobyan2019}
D.~Hakobyan and A.~Heuer, \emph{J. Chem. Theory Comput.}, 2019, \textbf{15},
  6393--6402\relax
\mciteBstWouldAddEndPuncttrue
\mciteSetBstMidEndSepPunct{\mcitedefaultmidpunct}
{\mcitedefaultendpunct}{\mcitedefaultseppunct}\relax
\EndOfBibitem
\end{mcitethebibliography}
\balance

\newpage

\begin{@twocolumnfalse}

\renewcommand{\thesection}{S\arabic{section}}
\renewcommand{\thetable}{S\arabic{table}}
\renewcommand{\thefigure}{S\arabic{figure}}
\setcounter{figure}{0}
\setcounter{table}{0}
\setcounter{page}{1}

\section*{ESI}

The following files are available free of charge. \\

	\begin{center}

	\includegraphics[width=0.6\textwidth]{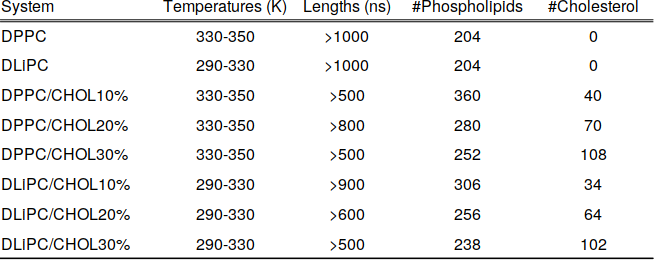}
	\captionof{table}{md\_parameters.png: Parameters of the MD simulations.}
	\label{tab:MDparam}

	\includegraphics[width=0.6\textwidth]{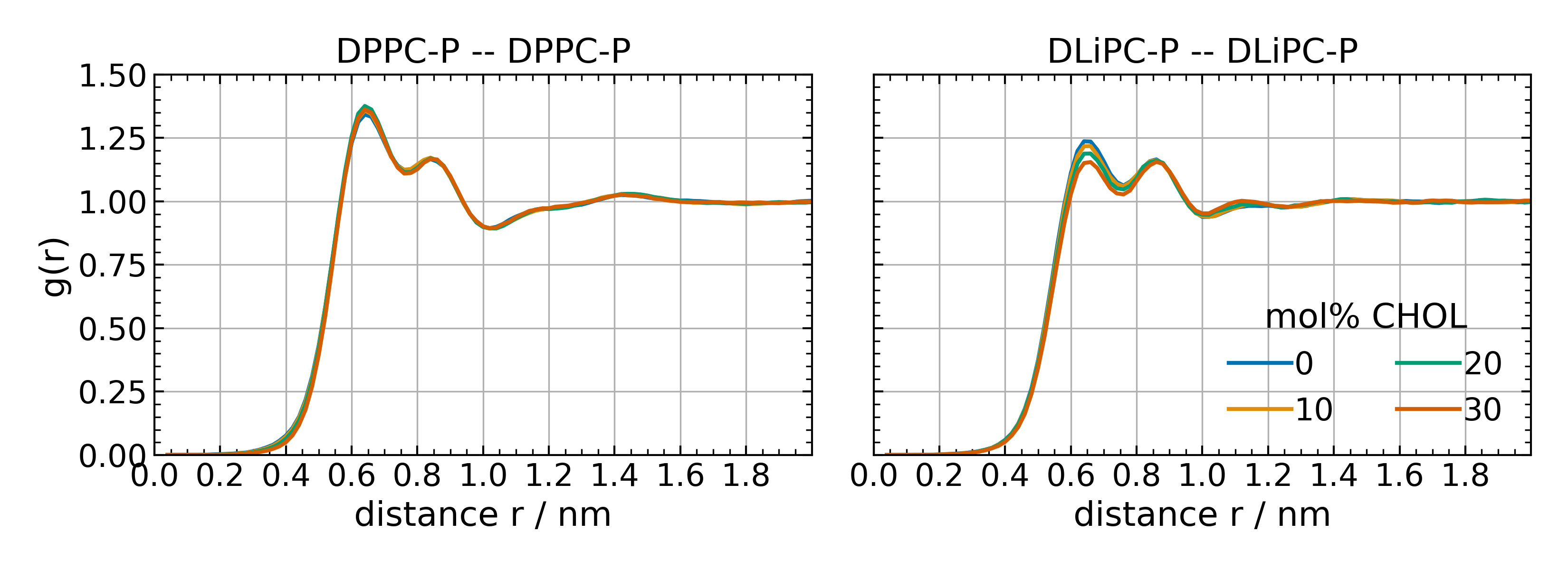}
	\captionof{figure}{gofr\_PL-PL.png: Radial distribution functions for lateral distances between P atoms. The RDFs were calculated individually for each leaflet and averaged.}
	\label{fig:rdf_PLPL}

	\includegraphics[width=0.6\textwidth]{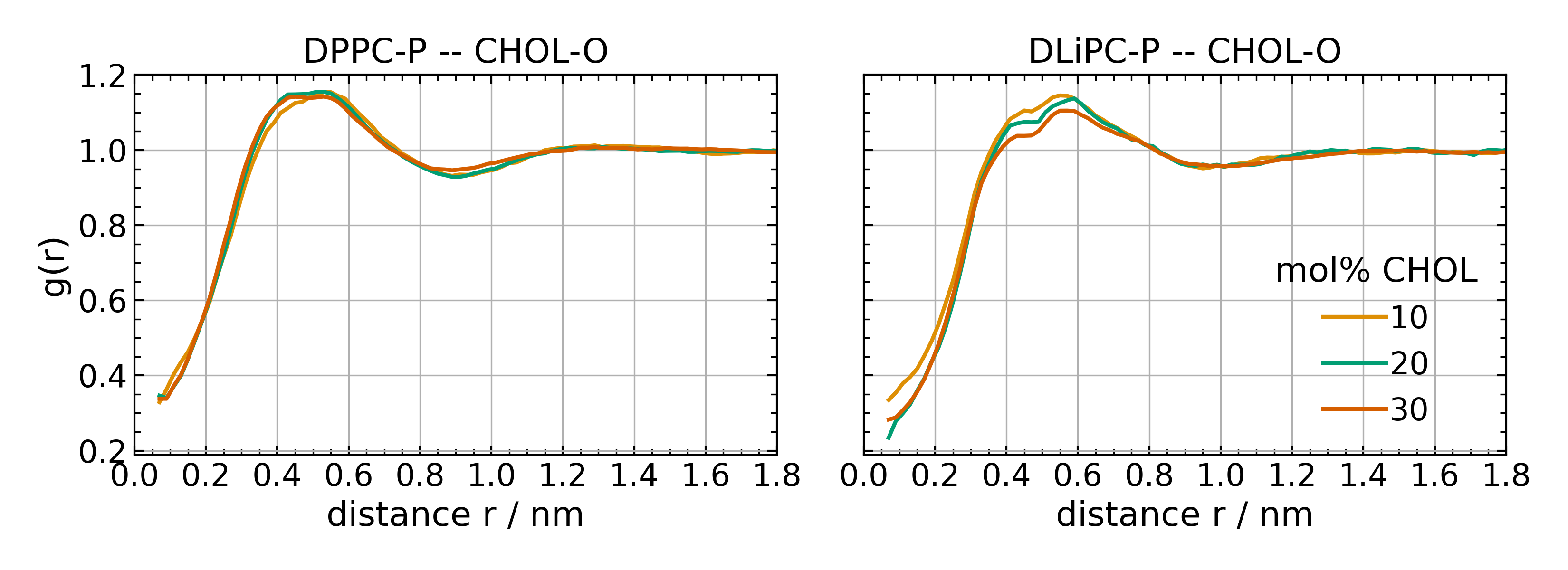}
	\captionof{figure}{gofr\_PL-O.png: Radial distribution functions for lateral distances between P and O atoms. The RDFs were calculated individually for each leaflet and averaged.}
	\label{fig:rdfPLO}

	\includegraphics[width=0.6\textwidth]{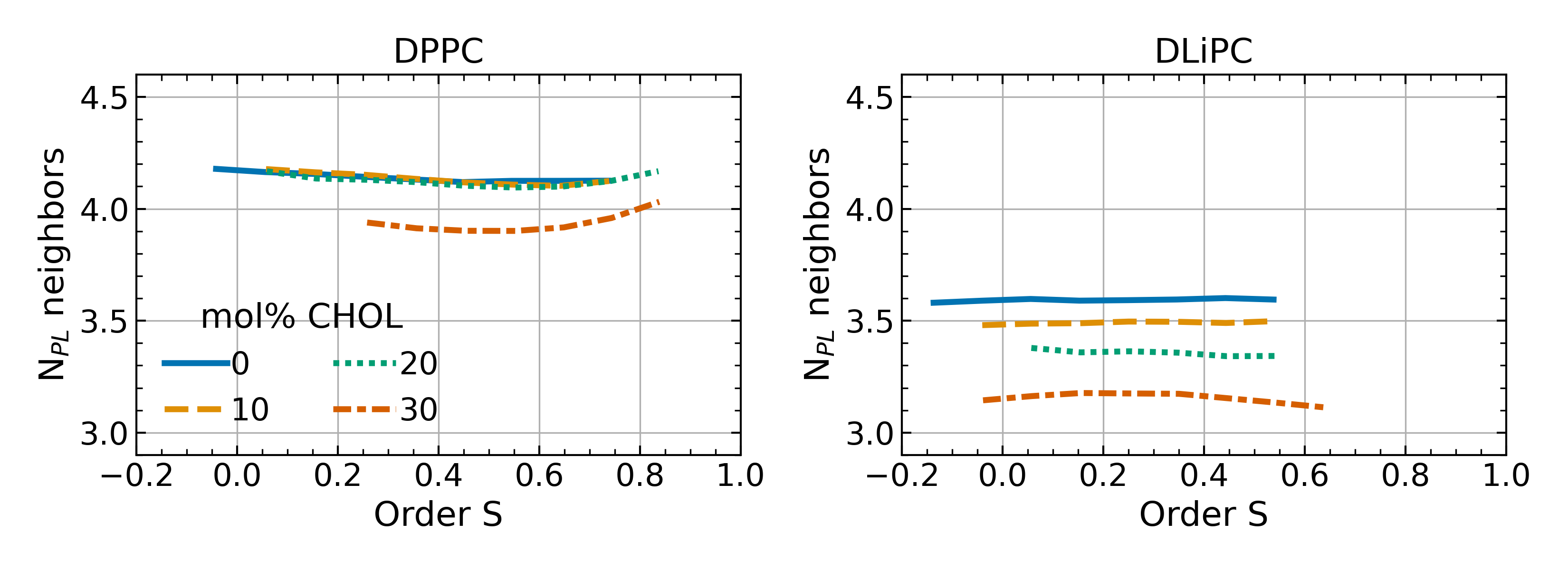}
	\captionof{figure}{Nneibs.png: Average number of nearest PL neighbors of DPPC (left) and DLiPC (right) as a function of the lipid's order parameter.}
	\label{fig:Nneibs}

	\includegraphics[width=0.6\textwidth]{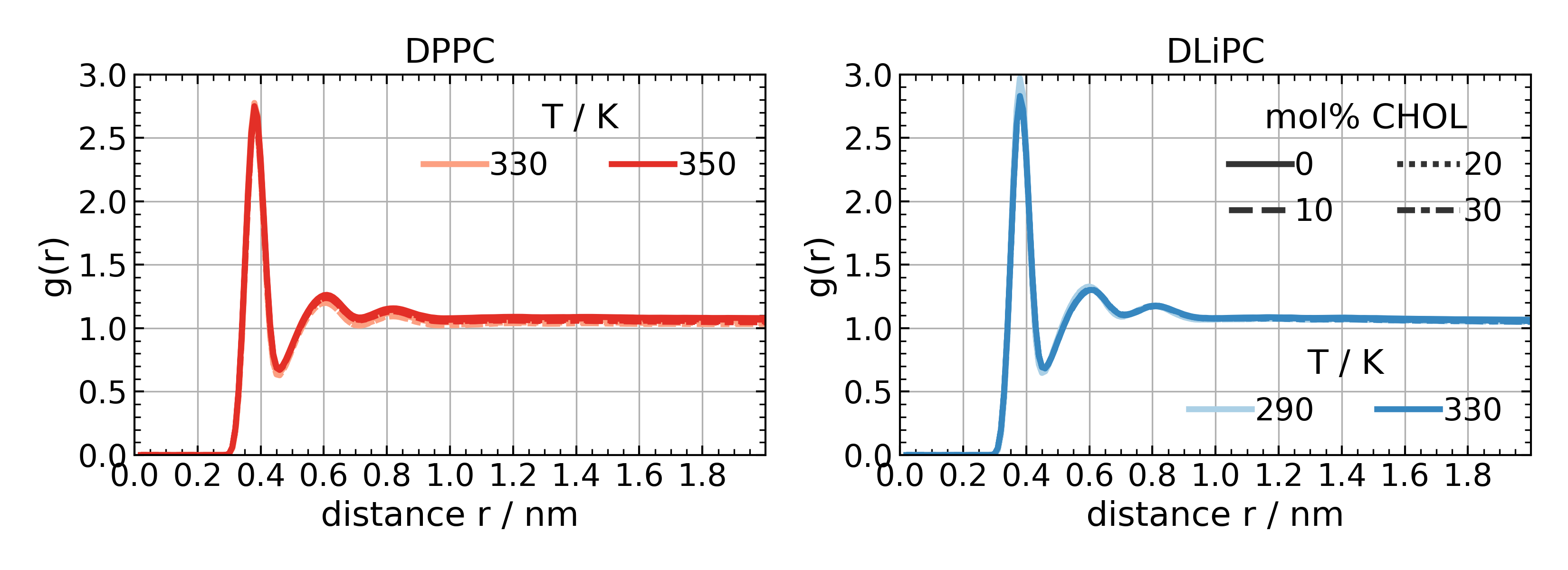}
	\captionof{figure}{rdf\_OH2.png: Radial distribution functions for distances between P and O atoms of surrounding water.}
	\label{fig:rdfh2o}

    \includegraphics[width=0.6\textwidth]{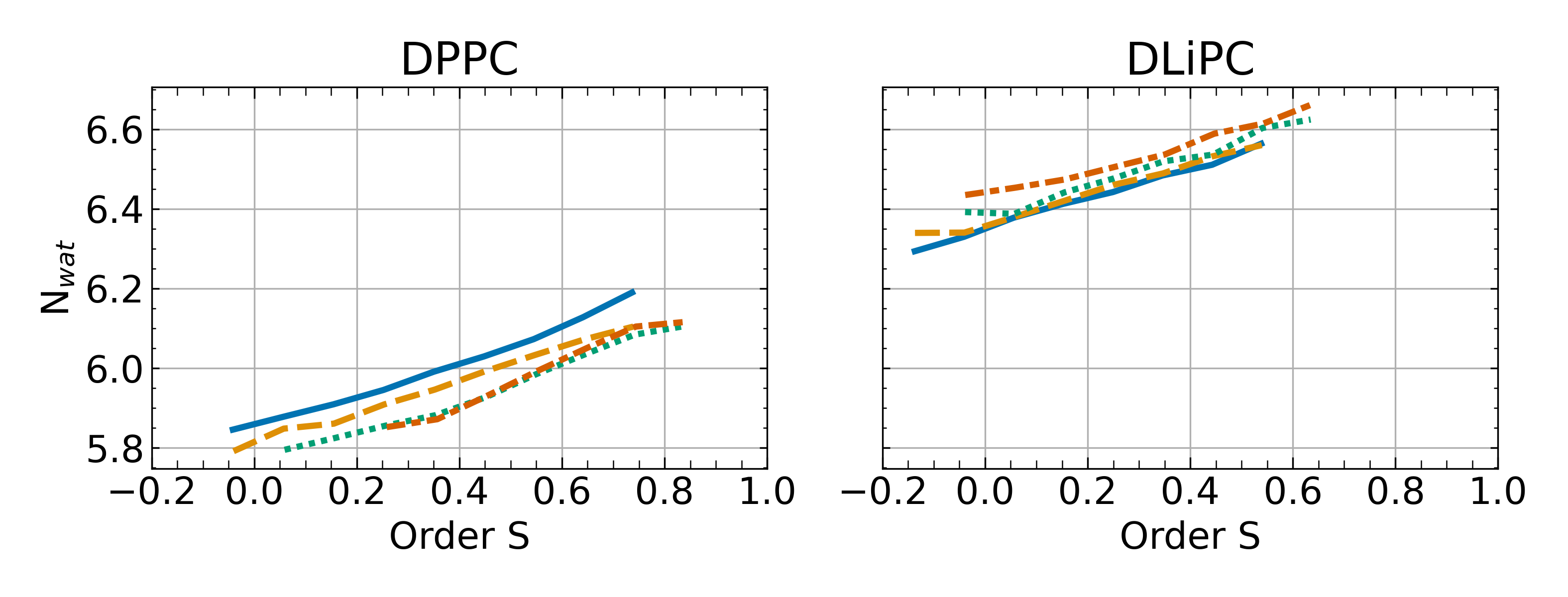}
    \captionof{figure}{watneibs.png: Average number of water molecules around PL phosphor atoms as a function of the PL chain order parameter. Solid blue, dashed yellow, dotted green and dot-dashed red indicate that energies were derived from bilayers with CHOL concentrations of 0, 10, 20 and 30 mol\% respectively.}
    \label{fig:watneibs}

    \includegraphics[width=0.6\textwidth]{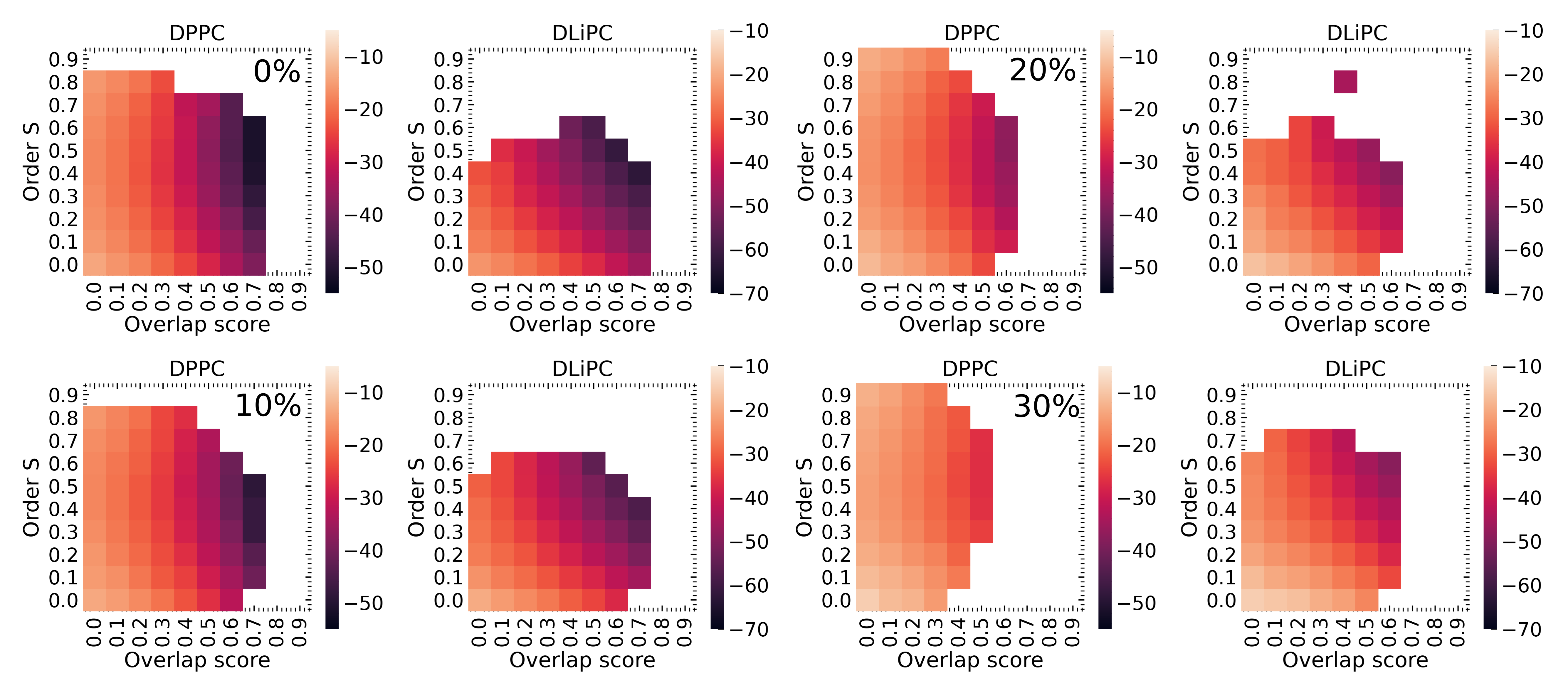}
    \captionof{figure}{2d\_overlap.png: Interleaflet interaction as a function of both, the lipids overlap and its chain order parameter in respective bilayer DPPC/CHOL and DLiPC/CHOL mixtures from 0\% to 30\% CHOL. }
    \label{fig:overlapmaps}
\pagebreak
    \includegraphics[width=0.3\textwidth]{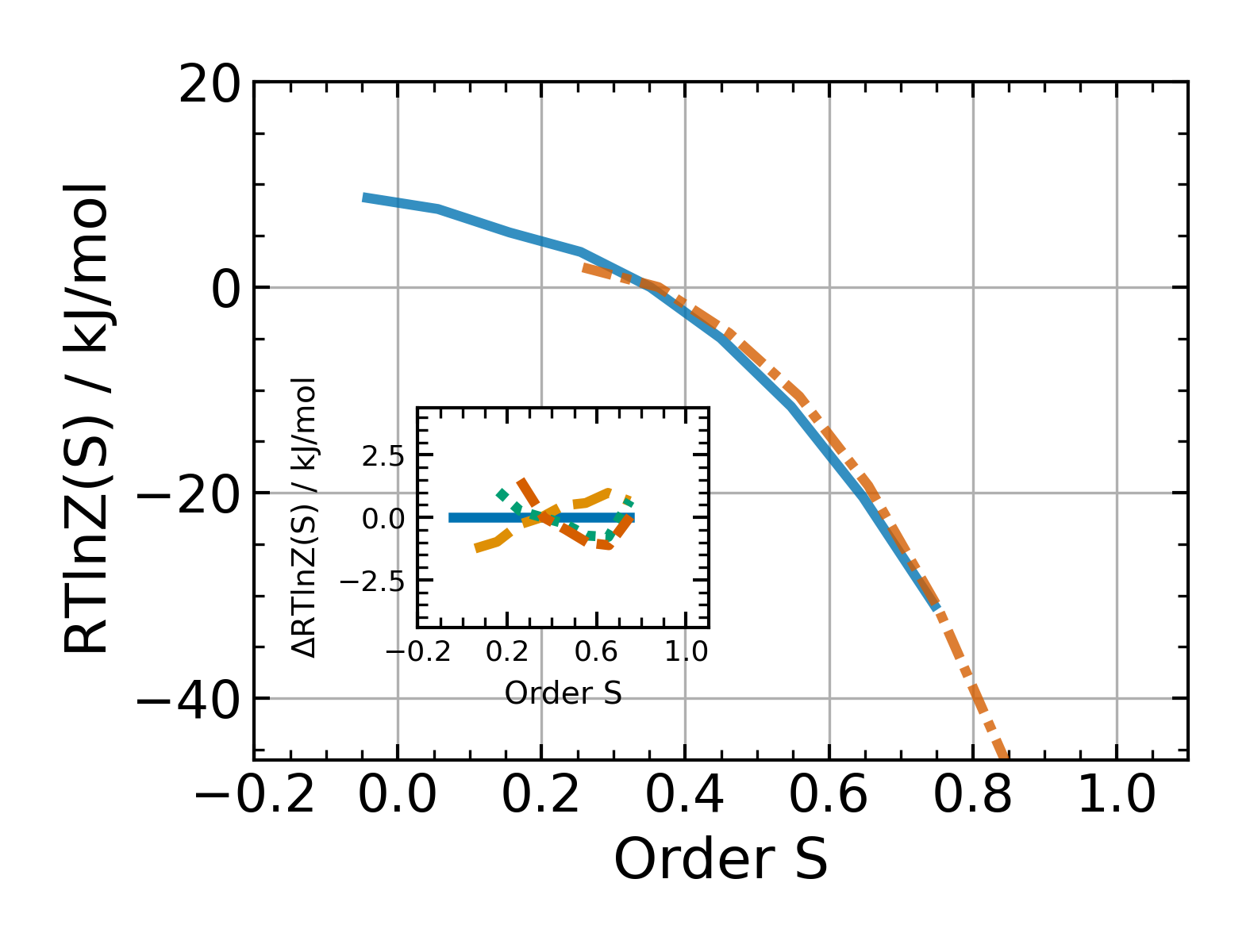}
    \captionof{figure}{zofs\_DPPC.png: Estimation of the acyl chain entropy of DPPC derived from the total sum of all enthalpic energy contributions and the order parameter distributions in the simulations at two CHOL concentrations and the difference between between the respective estimates in bilayer composition from 0\% to 30\% CHOL (inset). Solid blue, dashed yellow, dotted green and dot-dashed red indicate that energies were derived from bilayers with CHOL concentrations of 0, 10, 20 and 30 mol\% respectively.}
    \label{fig:zofs_DPPC}

    \includegraphics[width=0.3\textwidth]{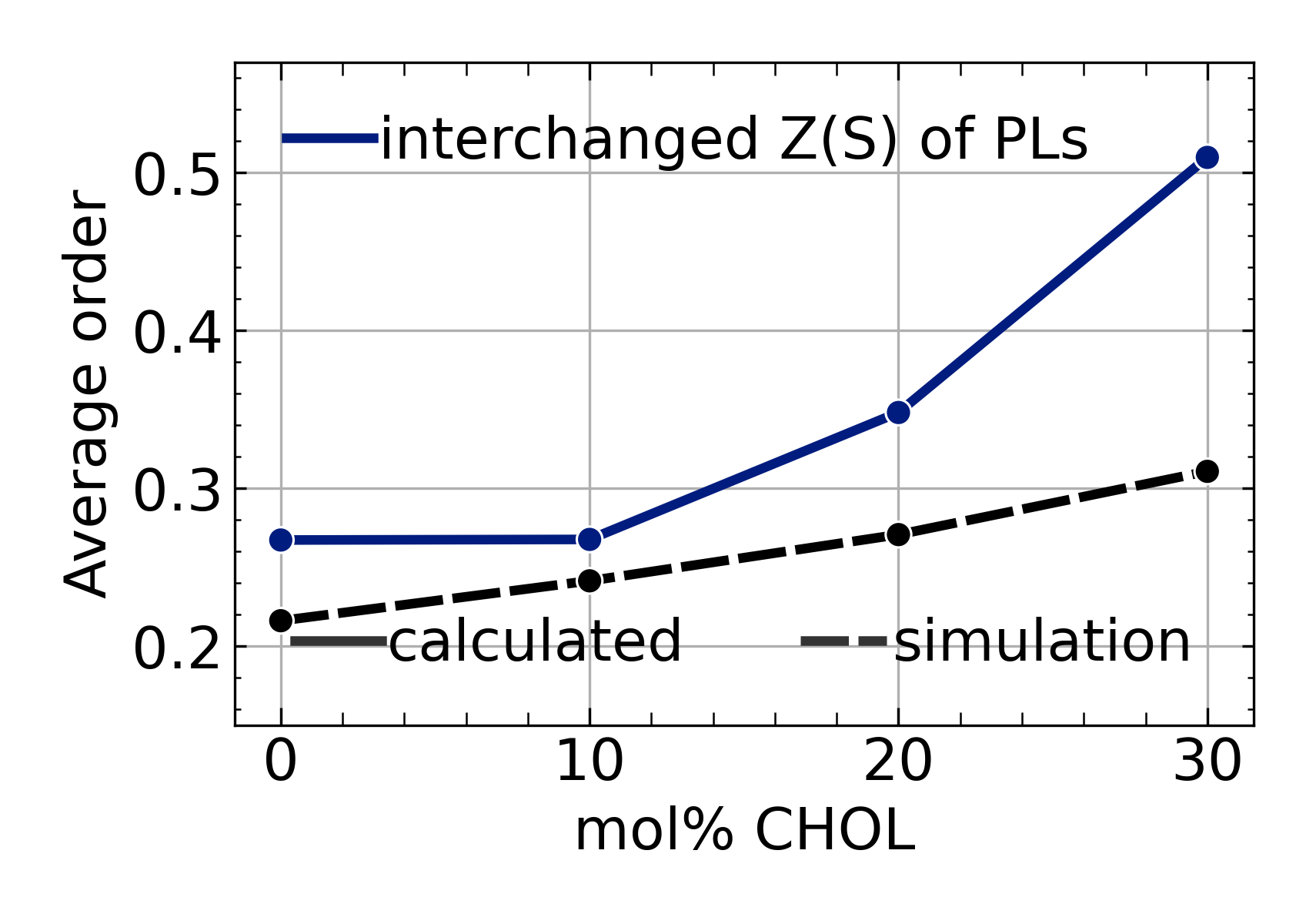}
    \captionof{figure}{avgs\_from\_zofs\_DLiPC\_Cdep.png: Average order parameter of DLiPC in a bilayer simulation at 330~K (dashed) as a function of CHOL concentration, the recalculated average order parameter when the $H_{DPPC}(S)$ is used along with $Z_{DLiPC}(S)$ (solid).}
    \label{fig:pfromz_DLiPC_Cdep}
\end{center}
\newpage

\end{@twocolumnfalse}

\end{document}